\documentclass[a4paper,10pt]{article}
\usepackage[english]{babel}
\usepackage{jheppub}
\usepackage[T1]{fontenc}
\usepackage{graphicx,calc}
\usepackage{epsfig}
\usepackage{amsmath}
\usepackage{amssymb}
\usepackage{float}
\usepackage{placeins}
\usepackage{braket}
\usepackage{slashed}
\usepackage{mathdots}
\usepackage{lipsum}
\usepackage{feynmf}
\usepackage{bbm}
\usepackage{color}
\usepackage{caption}
\usepackage{array}
\usepackage[export]{adjustbox}
\usepackage{subcaption}
\usepackage{multicol}
\usepackage{comment}

\allowdisplaybreaks[4]

\preprint{CERN-TH-2024-048, ZU-TH 22/24}

\title{Two-loop integrals for $t\bar t +$jet production at hadron colliders in the leading colour approximation}

\author[a]{Simon Badger}
\author[b]{Matteo Becchetti}
\author[c]{Nicol\`o Giraudo}
\author[d]{Simone Zoia}
\affiliation[a]{Physics Department, Torino University and INFN Torino, Via Pietro Giuria 1, I-10125 Torino, Italy}
\affiliation[b]{Dipartimento di Fisica e Astronomia, Università di Bologna e INFN, Sezione di Bologna, via Irnerio 46, I-40126 Bologna, Italy}
\affiliation[c]{Physik-Institut, Universität Zürich, Winterthurerstrasse 190, CH-8057 Zürich, Switzerland}
\affiliation[d]{CERN, Theoretical Physics Department, CH-1211 Geneva 23, Switzerland}
\emailAdd{simondavid.badger@unito.it}
\emailAdd{matteo.becchetti@unibo.it}
\emailAdd{nicolo.giraudo@physik.uzh.ch}
\emailAdd{simone.zoia@cern.ch}

\abstract{
We compute the differential equations for the two remaining integral
topologies contributing to the leading colour two-loop amplitudes for $pp
\rightarrow t\bar{t}j$. We derive differential equations for the master
integrals by solving the integration-by-parts identities over finite
fields. Of the two systems of differential equations, one is presented in canonical `${\rm d} \log$' form, while the
other is found to have an elliptic sector.
For the elliptic topology we identify the relevant elliptic curve, and present the differential equations in a more general form which depends quadratically on $\eps$ and contains non-logarithmic one-forms in addition to the canonical ${\rm d} \log$'s. 
We solve the systems
of differential equations numerically using generalised series expansions with the
boundary terms obtained using the auxiliary mass flow method. A summary of all
one-loop and two-loop planar topologies is presented including the list of
alphabet letters for the `${\rm d} \log$' form systems and high-precision boundary
values.
}

\newcommand{\beq}{\begin{equation}}

\newcommand{\eeq}{\end{equation}}

\newcommand{\bea}{\begin{eqnarray}}
\newcommand{\eea}{\end{eqnarray}}
\newcommand{\bfig}{\begin{figure}}
\newcommand{\efig}{\end{figure}}
\newcommand{\bc}{\begin{center}}
\newcommand{\ec}{\end{center}}

\newcommand{\eps}{{\varepsilon}}
\newcommand{\tb}{{\bar{t}}}

\newcommand{\cI}{{\mathcal{I}}}

\newcommand{\ii}{{\mathrm{i}}}
\newcommand{\dd}{{\mathrm{d}}}

\def\la{\langle}

\def\spAB#1#2#3{\la#1|#2|#3]}

\def\trfive{\operatorname{tr}_5}
\def\trp{\operatorname{tr}_+}
\def\trm{\operatorname{tr}_-}

\date{}
\begin{document}
\maketitle
\flushbottom

\section{Introduction}

Precision tests of the Standard Model (SM) in the top quark sector are essential part of the current physics program at high energy colliders. Observables
involving a top-quark pair have a wide variety of applications, and can be used to constrain SM parameters and parton distribution functions.
Furthermore, they are also important backgrounds for other searches.
Top-quark pair production in association with at least one hard jet forms a significant fraction,
around 50\%, of all top-quark pair events and has been studied extensively by both ATLAS and CMS
experiments~\cite{CMS:2016oae,ATLAS:2018acq,CMS:2020grm,CMS:2024ybg}.

The normalised differential cross section for the $t\tb j$ invariant mass in top-quark pair production in association with one jet ($pp\to t\tb j$) has
been proposed as an observable highly sensitive to the top-quark mass~\cite{Alioli:2013mxa}. Phenomenological studies and experimental analyses at
next-to-leading order (NLO) in QCD have already demonstrated this as a viable method for the extraction of the mass with current experimental
data~\cite{Bevilacqua:2017ipv,Alioli:2022lqo,ATLAS:2019guf}. While the current experimental and theoretical uncertainties are comparable, higher order
theoretical predictions will be necessary to make significant improvements as more data become available.

Precision theoretical predictions for top-quark processes present one of the most challenging classes of computation. The presence of internal massive
particles creates analytic complexity in the corresponding Feynman integrals, which means that conventional methods for their numerical evaluation can fail.
How to achieve a fast and stable evaluation of the integrals remains an active field of research, see~\cite{Bourjaily:2022bwx} and references therein. Coupling this with the
multiple scales associated with the three-parton final states makes processes such as $pp\to t\tb j$ and $pp\to t\tb W/Z/H$ seriously daunting computations.

The NLO QCD corrections have been available since 2007~\cite{Dittmaier:2007wz,Dittmaier:2008uj}. Remarkable theoretical developments since then now
enable a broad range of phenomenological predictions including full off-shell decays and interfaces with parton
shower~\cite{Melnikov:2010iu,Alioli:2011as,Czakon:2015cla,Bevilacqua:2015qha,Bevilacqua:2016jfk}, as well as mixed QCD and EW
corrections~\cite{Gutschow:2018tuk}. Improving on the theoretical precision requires going to next-to-next-to-leading order (NNLO) in QCD and, while the
tree-level and one-loop level amplitude ingredients can be provided by the current generation of automated tools~\cite{Campbell:2022qmc}, the two-loop
amplitudes remain a bottleneck. In this article, we continue the task of computing the two-loop QCD corrections to top-quark pair production in
association with a jet that was initiated in refs.~\cite{Badger:2022mrb, Badger:2022hno}. The preliminary tasks undertaken so far explored the
kinematic complexity of the helicity amplitudes at one-loop up to $\mathcal{O}(\eps^2)$ in the dimensional regularisation parameter $\eps$, and the
computation of master integrals relevant for a single `pentagon-box' two-loop integral family via canonical form differential
equations~\cite{Kotikov:1990kg,Remiddi:1997ny,Henn:2013pwa}.

The last few years have seen rapid progress for $2\to 3$ scattering amplitudes with massless internal propagators, in which a lot has been understood
for both the relevant Feynman integrals and the analytic representation of the amplitudes. Two new methods have had a major impact on these
developments: the widespread use of finite field modular arithmetic for algebraic
manipulations and rational function reconstruction~\cite{vonManteuffel:2014ixa,Peraro:2016wsq,Peraro:2019svx}, and the construction of special function bases, commonly referred to as
`pentagon functions', suitable for fast and stable numerical evaluations~\cite{Gehrmann:2015bfy,Gehrmann:2018yef,Chicherin:2020oor,Chicherin:2021dyp,Abreu:2023rco}.
Thanks to these developments, a plethora of new results have now been obtained for the one- and two-loop five-particle Feynman integrals relevant for the cases with no or one external massive particle~\cite{Gehrmann:2015bfy,Papadopoulos:2015jft,Abreu:2018aqd,Chicherin:2018old,Abreu:2020jxa,Canko:2020ylt,Abreu:2021smk,Abreu:2023rco}, scattering amplitudes~\cite{Gehrmann:2015bfy,Abreu:2018aqd,Badger:2018enw,Abreu:2018jgq,Chicherin:2018yne,Abreu:2018zmy,Badger:2019djh,Abreu:2019odu,Hartanto:2019uvl,Chicherin:2019xeg,Abreu:2020cwb,Chawdhry:2020for,Caron-Huot:2020vlo,Abreu:2021asb,Badger:2021ega,Agarwal:2021grm,Badger:2021imn,Chawdhry:2021mkw,Badger:2021nhg,Abreu:2021oya,Agarwal:2021vdh,Badger:2022ncb,Hartanto:2022qhh,Abreu:2023bdp,DeLaurentis:2023izi,Badger:2023mgf,DeLaurentis:2023nss,Agarwal:2023suw}, and
differential distributions at NNLO QCD accuracy~\cite{Chawdhry:2019bji,Kallweit:2020gcp,Chawdhry:2021hkp,Badger:2021ohm,Chen:2022ktf,Buonocore:2022pqq,Hartanto:2022qhh,Hartanto:2022ypo,Badger:2023mgf,Mazzitelli:2024ura}. The first results for massless six-particle kinematics have also recently appeared~\cite{Henn:2021cyv,Henn:2024ngj} together with the first steps investigating processes involving internal massive propagators~\cite{Badger:2022mrb,Badger:2022hno,FebresCordero:2023gjh,Buccioni:2023okz,Agarwal:2024jyq}.
Approximations of the two-loop amplitudes for $pp \to t\bar{t} H$ may also be used to make precise phenomenological predictions~\cite{Catani:2022mfv,Wang:2024pmv}.

This article presents the completion of the Feynman integral topologies needed for $pp\to t\tb j$ at two loops in the leading colour approximation.
The construction of the master integral bases and their evaluation follow the methodology used in the previous works~\cite{Badger:2022mrb,
Badger:2022hno}. Of the two new pentagon-box topologies that appear, one is shown to have some complicated features that go against the conventional
wisdom for what should appear for a leading colour, planar configuration. In particular, we find that one sector requires the use of a nested square root in order to
rotate the differential equations into an $\eps$-factorised form (see sec.~\ref{sec:421B}). This feature is perhaps not completely unexpected since
it has been observed before albeit with more complicated kinematics such as $pp\to t\tb H$~\cite{FebresCordero:2023gjh}. The second feature is that
one sector contains elliptic integrals (see sec.~\ref{sec:321B}). This is perhaps more surprising, as no sub-sectors of this topology were previously known to be elliptic, although
there are no principles forbidding it. 
Both of these features mean that the route towards a well defined basis of special functions which can can be
efficiently evaluated numerically is unclear. Canonical form differential equations have been studied in the case of elliptic Feynman integrals
although not for such complicated kinematics~\cite{Broedel:2017kkb,Broedel:2018qkq,Broedel:2018iwv,
Broedel:2019hyg,Frellesvig:2023iwr,Gorges:2023zgv}. Therefore in this article we present a compact representation of the differential
equations, which are not $\eps$-factorised, in terms of a minimal set of independent one-forms, and show that they can be reliably solved through generalised
series expansions~\cite{Francesco:2019yqt}.

The computation of the differential equations satisfied by the integral bases requires the solution of systems of
Integration-by-Parts (IBPs) identities~\cite{Tkachov:1981wb,Chetyrkin:1981qh}, which we generate using \textsc{LiteRed}~\cite{Lee:2012cn,Lee:2013mka}
and \textsc{NeatIBP}~\cite{Wu:2023upw}.  The latter package, in particular, allows us to obtain optimised systems of IBP relations through the
solution of syzygy equations~\cite{Gluza:2010ws}, this way making their solution substantially simpler.  We solve the IBP relations with the Laporta
algorithm~\cite{Laporta:2001dd} within the \textsc{FiniteFlow} framework~\cite{Peraro:2019svx,Peraro:2019okx}.

While we do not obtain canonical differential equations for the family involving elliptic integrals, we do put some effort into making their analytic
properties transparent.  In particular, we identify the elliptic curve associated with the appearing elliptic integrals.  Furthermore, we express the
differential equations in terms of a set of linearly independent one-forms.  We arrange the latter to be logarithmic as much as possible, in analogy
with the usual canonical differential equations.  Only the one-forms which relate master integrals involving either the elliptic curve or the nested
square root are non-logarithmic.  This makes the expression of the differential equations compact, and separates out clearly the features in common
with the standard canonical cases from the more complicated ones.  We expect that this information will be precious in view of future work to obtain a
canonical form for the differential equations of this topology as well.

We provide a semi-analytic solution to the differential equations for the master integrals by means of the multivariate generalisation of the method
of generalised power series expansions~\cite{Pozzorini:2005ff,Aglietti:2007as,Lee:2017qql,Lee:2018ojn,Bonciani:2018uvv,Fael:2021kyg,Fael:2022rgm}
proposed in ref.~\cite{Francesco:2019yqt} and implemented in a number of public \textsc{Mathematica}
packages~\cite{Hidding:2020ytt,Liu:2022chg,Armadillo:2022ugh}.  In particular, we make use of \textsc{DiffExp}~\cite{Hidding:2020ytt}.  The solution
to the differential equations is fully characterised once a set of boundary values is given.  We obtain high-precision numerical boundary values by
using the \textsc{Mathematica} package \textsc{AMFlow}~\cite{Liu:2022chg}, which implements the auxiliary mass flow
method~\cite{Liu:2017jxz,Liu:2021wks,Liu:2022tji}.  We use the interface to \textsc{FiniteFlow}~\citep{Peraro:2019svx} and
\textsc{LiteRed}~\cite{Lee:2012cn,Lee:2013mka} for the required IBP reduction.

The paper is structured as follows.  In section~\ref{sec:setup} we define the pentagon-box topologies under study and describe the computational
framework.  In section~\ref{sec:deqs} we discuss the construction and the analytic features of the master integral bases.  Section~\ref{sec:oneforms}
is devoted to the one-form representation of the differential equations, while in section \ref{sec:numerics} we discuss the numerical evaluation of
the master integrals.  We draw our conclusions and give an outlook on future developments in section~\ref{sec:conclusion}.  Finally, we identify the
elliptic curve underlying the analytic structure of the Feynman integrals in the elliptic sector in appendix~\ref{sec:CutBaikov}, define the one-loop
families in appendix~\ref{app:pentagons}, and describe the supplementary material in appendix~\ref{sec:anc}.

\section{Notation and definitions} \label{sec:setup}

\begin{figure}[t!]
	\centering
	\begin{subfigure}{0.35\linewidth}
		\includegraphics[width=\linewidth]{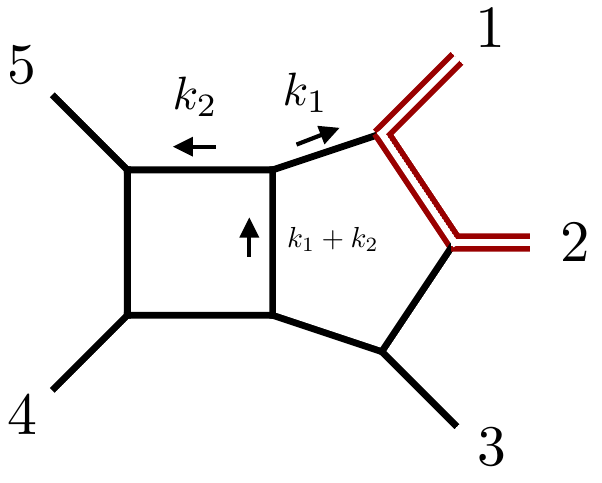}
		\caption{Topology ${\rm PB}_A$.}
		\label{fig:PBttjA}
	\end{subfigure}
	\begin{subfigure}{0.35\linewidth}
		\includegraphics[width=\linewidth]{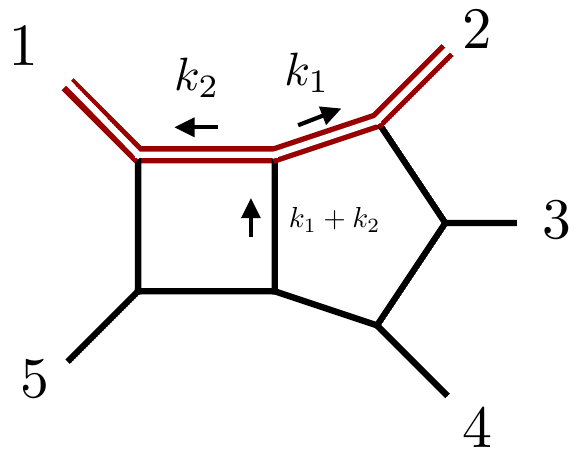}
		\caption{Topology ${\rm PB}_B$.}
		\label{fig:PBttjB}
	\end{subfigure} \\ \vspace{0.2cm}
	\begin{subfigure}{0.35\linewidth}
		\includegraphics[width=\linewidth]{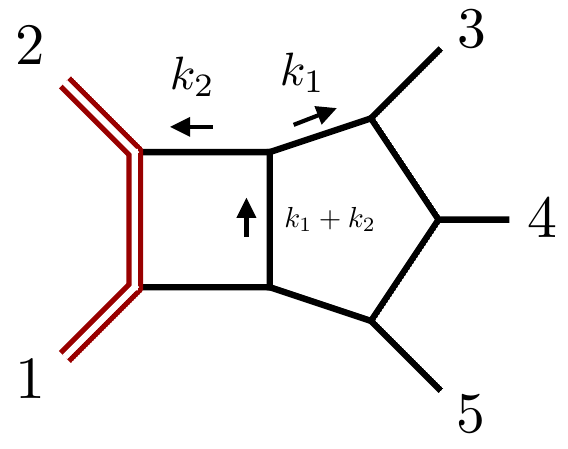}
		\caption{Topology ${\rm PB}_C$.}
		\label{fig:PBttjC}
	\end{subfigure}
	\caption{The three pentagon-box topologies contributing to $pp \rightarrow
t\bar{t}j$ in the leading colour limit. Black lines denote massless
particles and red double-lines denote massive particles.}
	\label{fig:topos}
\end{figure}

In this article we consider the three two-loop pentagon-box integral
topologies shown in figure~\ref{fig:topos}. In addition, we present an explicitly
`$\dd \log$' form for all one-loop pentagon integrals, which we define in appendix~\ref{app:pentagons}.
The integrals of the pentagon-box topology $F\in \{{\rm PB}_A, {\rm PB}_B, {\rm PB}_C\}$ have the form
\begin{equation}
  I_{a_1,a_2,a_3,a_4,a_5,a_6,a_7,a_8}^{\left(F\right),a_9,a_{10},a_{11}} = \int \mathcal{D}^{d} k_1 \, \mathcal{D}^{d} k_2 \,
  \frac{D_{F,9}^{a_9} D_{F,10}^{a_{10}} D_{F,11}^{a_{11}}}{D_{F,1}^{a_1}\cdots D_{F,8}^{a_8}}\,,
  \label{eq:t431def}
\end{equation}
where $a_1,\cdots,a_{11} \geq 0$ and $d=4-2\eps$. 
The integration measure is
\begin{equation} \label{eq:measure}
  \mathcal{D}^d k_i = \mathrm{e}^{\eps \gamma_{\rm E}} \dfrac{\dd^d k_i}{\ii \pi^{\frac{d}{2}}}  \,.
\end{equation}
The inverse propagators $D_{F,i}$ are defined in table~\ref{tab:PB-propagators}.
Note that $D_{F,9}$, $D_{F,10}$ and $D_{F,11}$ are irreducible scalar products. 
\begin{table}[h!]
\begin{center}
\begin{tabular}{|c|c|c|c|}
\hline
 & ${\rm PB}_A$ & ${\rm PB}_B$ & ${\rm PB}_C$ \\
\hline
$D_{F,1}$ & $k_1^2$ & $k_1^2 - m_t^2$ & $k_1^2$ \\
$D_{F,2}$ & $(k_1 - p_1)^2 -m_t^2$ & $ (k_1 - p_2)^2$ & $(k_1 - p_3)^2$ \\
$D_{F,3}$ & $(k_1 - p_1 - p_2)^2$ & $ (k_1 - p_2 - p_3)^2$ & $ (k_1 - p_3 - p_4)^2 $ \\
$D_{F,4}$ & $(k_1 - p_1 - p_2 - p_3)^2$ & $(k_1 - p_2 - p_3 - p_4)^2$ & $(k_1 - p_3 - p_4 - p_5)^2$ \\
$D_{F,5}$ & $k_2^2$ & $k_2^2 - m_t^2$ & $ k_2^2$ \\
$D_{F,6}$ & $(k_2 - p_5)^2$ & $(k_2 - p_1)^2$ & $ (k_2 - p_2)^2 - m_t^2$ \\
$D_{F,7}$ & $ (k_2 - p_4 - p_5)^2$ & $(k_2 - p_1 - p_5)^2$ & $(k_2 - p_1 - p_2)^2$ \\
$D_{F,8}$ & $(k_1 +k_2)^2$ & $ (k_1 +k_2)^2$ & $ (k_1 +k_2)^2$ \\
$D_{F,9}$ & $(k_1 + p_5)^2$ & $(k_1 + p_1)^2 - m_t^2$ & $ (k_1 + p_2)^2 - m_t^2$ \\
$D_{F,10}$ & $(k_2 + p_1)^2 - m_t^2$ & $(k_2 + p_2)^2 -m_t^2$ & $ (k_2 + p_3)^2$ \\
$D_{F,11}$ & $ (k_2 + p_1 + p_2)^2$ & $(k_2 + p_2 + p_3)^2 - m_t^2$ & $(k_2 + p_3 + p_4)^2$ \\
\hline
\end{tabular}
\end{center}
\caption{Inverse propagators $D_{F,i}$ of the pentagon-box topologies shown in figure~\ref{fig:topos}.}
\label{tab:PB-propagators}
\end{table}

The external momenta $p_i$ are considered as outgoing from the graphs and all the particles are on-shell, i.e.\ $p_1^2 = p_2^2 = m_t^2$ while $p_3^2 = p_4^2 =
p_5^2=0$. The kinematics of the integrals can be described in terms of six
independent Lorentz invariants. Here we choose the top-quark mass $m_t^2$ and the five adjacent scalar products, 
\begin{align}
\label{eq:x}
\vec{x}=\{d_{12}, d_{23}, d_{34}, d_{45}, d_{15}, m_t^2\} \,, 
\end{align}
where
\begin{equation}
  d_{ij} = p_i \cdot p_j \,.
  \label{eq:dijdef}
\end{equation}

The minimal set of master integrals (MIs) is obtained by Integration-by-Parts
(IBP) reduction~\cite{Chetyrkin:1981qh,Chetyrkin:1979bj}. We generate the
systems of IBPs with the software
\textsc{LiteRed}~\cite{Lee:2012cn,Lee:2013mka} and
\textsc{NeatIBP}~\cite{Wu:2023upw}, and solve them via the Laporta
algorithm~\cite{Laporta:2000dsw} within the finite-field framework
\textsc{FiniteFlow}~\citep{Peraro:2019svx}. Topology ${\rm PB}_A$ has been considered
previously in ref.~\cite{Badger:2022hno} and has 88 MIs.  For the new cases
presented here we find 121 MIs for topology ${\rm PB}_B$ and 109 for topology
${\rm PB}_C$.\footnote{\textsc{NeatIBP} finds additional symmetry relations with respect
to \textsc{LiteRed}, reducing the number of MIs by 2 for each topology ${\rm PB}_B$ and
${\rm PB}_C$. The missing relations are a result of both particles 1 and 2 having the
same mass.} 

We find a set of 14 square roots appearing in the differential equations and
associated alphabet for all one-loop and planar two-loop topologies. These are
defined as
\begin{multicols}{2}
\noindent
\begin{align}
  \beta_{12} & = \sqrt{\frac{\operatorname{det}G(p_1,p_2)}{s_{12}}} = \sqrt{1-\frac{4 m_t^2}{s_{12}}} \,, \\
  \beta_{34} & = \sqrt{1-\frac{4 m_t^2}{s_{34}}} \,, \\
  \beta_{45} & = \sqrt{1-\frac{4 m_t^2}{s_{45}}} \,, \\
  \Delta_{3,1} &= \sqrt{-\operatorname{det}G(p_{1},p_{23})} \,, \\
  \Delta_{3,2} &= \sqrt{-\operatorname{det}G(p_{2},p_{15})} \,, \\
  \Delta_{3,3} &= \sqrt{-\operatorname{det}G(p_{1},p_{25})} \,, \\
  \Delta_{3,4} &= \sqrt{-\operatorname{det}G(p_{2},p_{13})} \,, \\
  \Lambda_1 & = \sqrt{\frac{\operatorname{det}Y^{0mm0}_{p_1|p_3|p_2|p_{45}}}{d_{23}^2 d_{13}^2}}  \,, \\
  \Lambda_2 & = \sqrt{\frac{\operatorname{det}Y^{0mmm}_{p_2|p_3|p_4|p_{51}}}{d_{23}^2 d_{34}^2}} \,,
\end{align}
\columnbreak
\begin{align}
  \Lambda_3 & = \sqrt{\frac{\operatorname{det}Y^{m0mm}_{p_{13}|p_2|p_4|p_5}}{d_{24}^2 d_{45}^2}} \,, \\
  \Lambda_4 & = \sqrt{\frac{\operatorname{det}Y^{m0mm}_{p_1|p_{23}|p_4|p_5}}{d_{15}^2 d_{45}^2}} \,, \\
  \Lambda_5 & = \sqrt{\frac{\operatorname{det}Y^{mmmm}_{p_{12}|p_3|p_4|p_5}}{d_{34}^2 d_{45}^2}} \,, \\
  \Lambda_6 &= \sqrt{(s_{15}-s_{23})^2 + \frac{4 s_{34} s_{45} m_t^2}{s_{12}}} \,, \\
  \trfive &= 4 \sqrt{\operatorname{det}G(p_3,p_4,p_5,p_1)} \,,
  \label{eq:tr5}
\end{align}
\end{multicols}
\noindent 
where $G_{ij}(\vec{v}) = v_i\cdot v_j $ is the Gram matrix, $p_{ij} = p_i+p_j$,
$s_{ij} = (p_i+p_j)^2$, and $m$ denotes $m_t$ for the sake of compactness. $Y^{m_1 \ldots m_4}_{P_1|\ldots|P_4}$ represents the
Cayley matrices associated with one-loop box configurations appearing at
sub-leading colour, which are defined by
\begin{align}
    \left [Y_{P_1|P_2|P_3|P_4}^{m_1 m_2 m_3 m_4} \right]_{ij} = \frac{1}{2} 
    \left[ (q_i-q_j)^2-m_i^2-m_j^2 \right] \,,\qquad
    q_i = \sum_{k=0}^{i-1} P_k \,,
  \label{eq:Cayley}
\end{align}
for $i,j=1,\ldots,4$, with $P_0 = 0$.
$\Lambda_6$ only appears at two loops.

Note that $\trfive$ is related to the five-point pseudo-scalar invariant via
\begin{align}
\trfive^2 = {\rm tr}(\gamma_5 \slashed{p}_3 \slashed{p}_4 \slashed{p}_5 \slashed{p}_1)^2 \,.
\end{align}
We refrain from identifying $\trfive$ with ${\rm tr}(\gamma_5 \slashed{p}_3 \slashed{p}_4 \slashed{p}_5 \slashed{p}_1)$, as the latter is a parity-odd object, i.e.\ it changes sign under space-time parity conjugation and odd-signature permutations. 
The parity degree of freedom is required to describe scattering amplitudes, but
not for the computation of Feynman integrals.  We therefore prefer to define
$\trfive$ as the parity-even square root in eq.~\eqref{eq:tr5}.  Furthermore,
we will make use of the short-hand
\begin{align}
{\rm tr}_{\pm} \left[ \slashed{p}_i \slashed{p}_j \slashed{p}_k \slashed{p}_l \right] = \frac{1}{2} \, {\rm tr}\left[\slashed{p}_i \slashed{p}_j \slashed{p}_k \slashed{p}_l (1 \pm \gamma_5) \right] 
\end{align}
to make certain expressions more compact.
We however remove the parity degree of freedom in them by replacing the parity-odd
trace ${\rm tr}(\gamma_5 \slashed{p}_3 \slashed{p}_4 \slashed{p}_5
\slashed{p}_1)$ with the parity-even square root $\trfive$ defined in
eq.~\eqref{eq:tr5}.

In order to write compact expressions for the MIs which satisfy canonical
differential equations, we use choices motivated by the local numerators
introduced in
refs.~\cite{Arkani-Hamed:2010zjl,Arkani-Hamed:2010pyv,Gehrmann:2015bfy,Badger:2016ozq,Abreu:2020jxa}.
In $d=4-2\eps$ dimensions, we can find numerator insertions that can be written
in terms of the $-2\eps$ dimensional components of the loop momenta and are
conventionally denoted $\mu_{ij}$,
\begin{align}
  k_i &= k_i^{[4]} + k_i^{[-2\eps]}, & \mu_{ij} &= -k_i^{[-2\epsilon]}\cdot k_j^{[-2\epsilon]}.
\end{align}
We denote these numerator insertions using an additional superscript $[ij]$, as
\begin{align}
\begin{aligned}
  I_{a_1,a_2,a_3,a_4,a_5,a_6,a_7,a_8}^{(F),[ij],a_9,a_{10},a_{11}} & = \int \mathcal{D}^{d} k_1 \,  \mathcal{D}^{d} k_2 \, \mu_{ij} \,
   \frac{D_{F,9}^{a_9} D_{F,10}^{a_{10}} D_{F,11}^{a_{11}}}{D_{F,1}^{a_1}\cdots D_{F,8}^{a_8}}\, , \\
  I_{a_1,a_2,a_3,a_4,a_5,a_6,a_7,a_8}^{(F),[ij,kl],a_9,a_{10},a_{11}} & = \int \mathcal{D}^{d} k_1 \,  \mathcal{D}^{d} k_2 \, \mu_{ij} \, \mu_{kl} \,
   \frac{D_{F,9}^{a_9} D_{F,10}^{a_{10}} D_{F,11}^{a_{11}}}{D_{F,1}^{a_1}\cdots D_{F,8}^{a_8}}\,.
\end{aligned}
\end{align}

\section{Master integral bases} 
\label{sec:deqs}

In this section we discuss the construction of the bases of master integrals for topology ${\rm PB}_B$ and ${\rm PB}_C$, and describe their features.
The guiding principle in this construction is the simplification of the differential equations (DEs) satisfied by the MIs.
Let $\vec{\mathcal{I}}_{F}$ be the list of MIs for topology $F$. 
In general, $\vec{\mathcal{I}}_{F}$ satisfies a system of DEs of the form~\cite{Barucchi:1973zm,Kotikov:1990kg,Kotikov:1991hm,Gehrmann:1999as,Bern:1993kr}
\begin{align} \label{eq:DEsGeneral0}
\frac{\partial}{\partial x_i} \, \vec{\mathcal{I}}_{F}(\vec{x}, \eps) = A_{x_i}^{(F)}\left(\vec{x},\eps\right) \cdot \vec{\mathcal{I}}_{F}(\vec{x}, \eps)\,,
\end{align}
for every $i=1,\ldots,6$.
The matrices $A_{x_i}^{(F)}$ are called \emph{connection matrices}. 
We rewrite eq.~\eqref{eq:DEsGeneral0} in a more compact form by introducing the total differential with respect to the kinematic invariants, $\dd$, as
\begin{align}
\label{eq:DEsGeneral}
\dd \, \vec{\mathcal{I}}_{F}(\vec{x}, \eps) = \dd A^{(F)}\left(\vec{x},\eps\right) \cdot \vec{\mathcal{I}}_{F}(\vec{x}, \eps)\,,
\end{align}
where $\dd A^{(F)}$ is the matrix-valued one-form 
\begin{align}
\dd A^{(F)}(\vec{x}, \eps) = \sum_{i=1}^6 A_{x_i}^{(F)}\left(\vec{x},\eps\right) \, \dd x_i \,.
\end{align}
With a slight abuse of notation, we refer to $\dd A^{(F)}(\vec{x}, \eps)$ as connection matrix as well.
The solution to eq.~\eqref{eq:DEsGeneral} is enormously simplified if a choice of MIs is found such that the DEs take the \emph{canonical form}~\cite{Henn:2013pwa}
\begin{equation} \label{eq:DEsCanonical}
 \dd\, \vec{\mathcal{I}}_{F}(\vec{x},\eps) = \eps \, \dd A^{(F)}(\vec{x}) \, \vec{\mathcal{I}}_{F}(\vec{x},\eps) \,,
\end{equation} 
where the dependence of the connection matrices on $\eps$ is factorised, and $\dd A^{(F)}(\vec{x})$ is a linear combination of logarithmic one-forms:
\begin{equation}\label{eq:dAdlog}
\dd A^{(F)}(\vec{x}) = \sum_i c^{(F)}_i  \dd \log\left(W_i (\vec{x}) \right) \,.
\end{equation}
Here, the $c^{(F)}_i$ are matrices of rational numbers, and the \emph{letters} $W_i (\vec{x})$ are algebraic functions of the kinematic invariants $\vec{x}$.
Their ensemble, called \emph{alphabet}, dictates the singularity structure of the MIs. 
The factorisation of $\eps$ allows us to express the solution algorithmically in terms of Chen iterated integrals~\cite{Chen:1977oja}, order by order in the Laurent expansion around $\eps=0$.
This, in conjunction with the presence of logarithmic one-forms only, enables the application of a well-established toolbox of mathematical techniques ---~most notably the symbol~\cite{Goncharov:2010jf}~--- to write down and manipulate the solution.
Building on this, the method of the so-called \emph{pentagon functions}~\cite{Gehrmann:2018yef,Chicherin:2020oor,Badger:2021nhg,Chicherin:2021dyp,Badger:2023xtl,Abreu:2023rco,FebresCordero:2023gjh} has proven particularly successful in the computation of two-loop amplitudes for $2 \to 3$ processes.

It is however known that the DEs for Feynman integrals can take more complicated forms.
Indeed, we anticipate that the canonical form in eqs.~\eqref{eq:DEsCanonical} and~\eqref{eq:dAdlog} can only be achieved for topologies ${\rm PB}_A$ and ${\rm PB}_C$, whereas a generalisation is necessary for topology ${\rm PB}_B$.
First of all, it is not proven that one can always factorise the dependence on $\eps$ in the connection matrices. 
On top of that, even when $\eps$ is factorised, one-forms other than $\dd \log$'s may be necessary (see e.g.\ the review~\cite{Bourjaily:2022bwx} and references therein).
For such cases, the notion of `canonical' DEs is still under debate~\cite{Broedel:2018qkq,Frellesvig:2023iwr}. 
On the one hand, the techniques for bringing the DEs to an $\eps$-factorised form are much less mature than in the $\dd \log$ case.
Moreover, even when an $\eps$-factorised form is achieved which involves one-forms more complicated than the $\dd \log$'s in eq.~\eqref{eq:dAdlog}, manipulating and evaluating the solution efficiently remain challenging. 
For these reasons, it is often convenient to resort to more flexible numerical approaches to solve DEs beyond the $\dd \log$ case, as opposed to fully analytic solutions in terms of well understood special functions.
The method of generalised power series expansions~\cite{Francesco:2019yqt} is proving particularly effective, boosted by the availability of public implementations~\cite{Hidding:2020ytt,Liu:2022chg,Armadillo:2022ugh}.
Nonetheless, also within this approach to the solution, simplifying as much as possible the form of the DEs is crucial to an efficient and stable evaluation of the solution.
In particular, it is desirable for the connection matrices to depend polynomially on $\eps$, and for the degree in $\eps$ to be as low as possible. 
More explicitly, in the generalisation of the canonical DEs we consider in this work for topology ${\rm PB}_B$, the connection matrix has the form 
\begin{equation} \label{eq:DEsPolyEps}
\dd A^{({\rm PB}_B)}(\vec{x}, \eps) = \sum_{k=0}^{k_{\rm max}} \sum_i  \eps^k \, c^{({\rm PB}_B)}_{k,i} \, \omega_i(\vec{x}) \,,
\end{equation} 
where $\omega_i(\vec{x})$ are (linearly independent) \emph{one-forms}, $k_{\rm max} \in \mathbb{N}$, and $c^{({\rm PB}_B)}_{k,i}$ are matrices of rational numbers.
More explicitly, the one-forms have the form
\begin{align} \label{eq:one-forms}
\omega_i(\vec{x}) = \mathcal{C}_i(\vec{x}) \sum_{j=1}^6 f_{ij}(\vec{x}) \, \dd x_j \,,
\end{align}
where $f_{ij}(\vec{x})$ are rational functions, and $\mathcal{C}_i(\vec{x})$ is either $1$ or a square root (possibly a product of square roots).
A subset of the one-forms $\{\omega_i(\vec{x})\}$ may be logarithmic, i.e., for some one-form $\omega(\vec{x}) $ there may exist $W(\vec{x})$ such that $\omega(\vec{x}) = \dd \log W(\vec{x})$.
In this case, we recall that $W(\vec{x})$ is called \emph{letter}. 

Equation~\eqref{eq:one-forms} implies that each one-form (including the logarithmic ones) possesses a property called \emph{charge} with respect to the square roots of the problem. With respect to each square root $\mathcal{C}$, a one-form $\omega$ is either \emph{even}, if it stays invariant when we flip the sign of $\mathcal{C}$, or \emph{odd}, if it changes sign:
\begin{align}
\omega \bigl|_{\mathcal{C} \to - \mathcal{C}} = \begin{cases} \omega \,, \quad & \text{if $\omega$ is \emph{even} w.r.t.\ ${\cal C}$} \,, \\ - \omega\,, \quad & \text{if $\omega$ is \emph{odd} w.r.t.\ ${\cal C}$} \,.
\end{cases}
\end{align}
For a logarithmic one-form $\dd \log W$, this definition implies that the letter $W$ is even (odd) with respect to a square root $\mathcal{C}$ if $W\bigl|_{\mathcal{C} \to -\mathcal{C}} = W$ ($W\bigl|_{\mathcal{C} \to -\mathcal{C}} = 1/W$). For example, a typical odd letter has the form
\begin{align}
\frac{A+\mathcal{C}}{A-\mathcal{C}}\,,
\end{align}
where $\mathcal{C}$ is square root and $A$ is a rational function.
The fact that all one-forms appearing in the DEs for the MIs have a well-defined transformation under the change of the sign of the square roots follows from the way the square roots enter in the definition of the MIs.
The scalar integrals, i.e.\ the integrals of the form $I^{(F),a_9,a_{10},a_{11}}_{a_1,\ldots,a_8}$, are by definition even with respect to all square roots.
The square roots enter the game in the construction of MIs which satisfy DEs in the canonical form.
As we will see later in this section (e.g.\ see eq.~\eqref{eq:PBB-topsec}), the square roots appear as overall normalisation of the MIs.
More explicitly, each MI $\mathcal{I}$ has the form
\begin{align}
\mathcal{I}(\vec{x},\eps) = \mathcal{C}(\vec{x}) \, \sum_{i} c_i(\vec{x},\eps) \, I_i(\vec{x},\eps) \,,
\end{align}
where $ \mathcal{C}(\vec{x})$ is either $1$ or a product of square roots, $c_i(\vec{x},\eps)$ are rational functions, and $I_i$ are scalar integrals.
This way, the MIs gain a charge with respect to the square roots, and the entries of the connection matrices ---~and thus the one-forms~--- inherit it from the MIs.
With a slight abuse of notation, we say that a MI or one-form has charge $\mathcal{C}$ if it is odd with respect to $\mathcal{C}$.

\begin{figure}
	\centering
	\begin{subfigure}{0.35\linewidth}
		\includegraphics[width=\columnwidth]{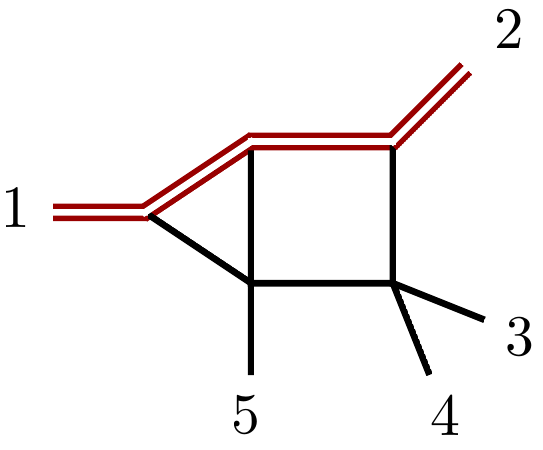}
		\caption{The elliptic sector which we refer to as $321B$.}
		\label{fig:321B}
	\end{subfigure}
	\hspace{1cm}
	\begin{subfigure}{0.35\linewidth}
		\includegraphics[width=\columnwidth]{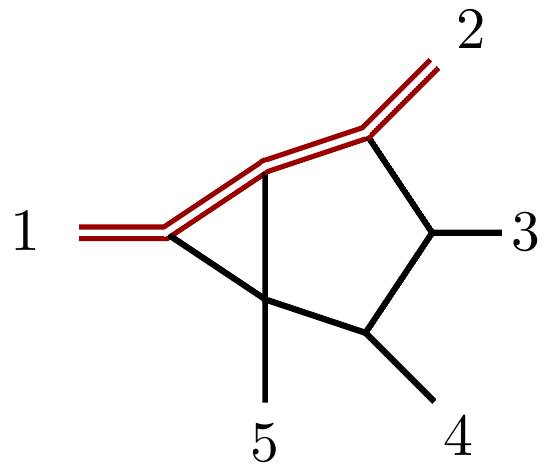}
		\caption{Pentagon-triangle sector containing a nested square root. We refer to it as $421B$}
		\label{fig:421B}
	\end{subfigure}
	\caption{`Problematic' sectors of topology ${\rm PB}_B$.}
	\label{fig:probSecB}
\end{figure}

In order to construct a basis of MIs which satisfies DEs of the previous forms, we used the approach outlined in ref.~\cite{Badger:2022mrb}.
For completeness, we briefly summarise it in the following.  It is a bottom-up approach, i.e.\ we start from the integral sectors with the fewest number
of propagators, and we bring the DEs to the desired form sector by sector. The simplification of the DEs is done by following a procedure which exploits a set
of heuristic criteria and draws as much as possible from known results in the literature. In particular, we perform the following steps.
Let $S$ be a sector of topology $F$.
\begin{itemize}
\item \textbf{Step 1}: We choose candidate MIs $ \vec{\mathcal{L}}_{F,S}(\vec{x},\epsilon)$ for sector $S$ by requiring that the homogeneous DE\footnote{Given an integral sector, the homogeneous DE
is the subset of a DE which contains the contributions coming only from the integrals of that sector. The corresponding connection matrix is zero everywhere except for the diagonal square block corresponding to the MIs of the chosen sector. As a consequence, since all the sub-topologies do not contribute to it, the maximal cut of a certain MI is also a solution to the homogeneous DEs of the corresponding sector~\cite{Primo:2016ebd}.} has the $\epsilon$ structure
\begin{equation}
\dd\, \vec{\mathcal{L}}_{F,S}(\vec{x},\epsilon) = \sum_{k=0}^{k_{\rm max}} \epsilon^k  \dd \tilde{A}^{(F,S)}_{k,\operatorname{Hom}}(\vec{x}) \cdot \vec{\mathcal{L}}_{F,S}(\vec{x},\epsilon) + (\text{sub-sectors}) \,,
\end{equation}  
where $k_{\rm max}=2$ for the elliptic sector in ${\rm PB}_B$, otherwise $k_{\rm max} =1$.
We neglect all contributions from the sub-sectors, which constitute the inhomogeneous terms of the DEs.
As a guiding principle, we select MI candidates following patterns observed in previously studied cases. 
Since in this step we are mostly interested in the $\epsilon$ structure of DEs for the candidate MIs under study, we exploit finite fields technique to perform IBP reduction and to reconstruct 
the DEs on a univariate $\epsilon$-slice, i.e.\ we set to numbers all the kinematic invariants and reconstruct just the analytic dependence in $\epsilon$.
\item \textbf{Step 2}: We reconstruct analytically the homogeneous DEs.
Except for the two problematic sectors shown in fig.~\ref{fig:probSecB}, which we will discuss later, 
we construct a rational transformation such that the DEs take the intermediate form
\begin{equation}
\dd\, \vec{\mathcal{J}}_{F,S}(\vec{x},\epsilon) = \left(\dd \hat{A}^{(F,S)}_{0,\operatorname{Hom}}(\vec{x}) + \epsilon \, \dd \hat{A}^{(F,S)}_{1,\operatorname{Hom}}(\vec{x})\right) \cdot \vec{\mathcal{J}}_{F,S}(\vec{x},\epsilon) + (\text{sub-sectors})\,,
\end{equation}  
where $\hat{A}^{(F,S)}_{0,\operatorname{Hom}}$, and $\hat{A}^{(F,S)}_{1,\operatorname{Hom}}$ are matrices of rational functions, and $\hat{A}^{(F,S)}_{0,\operatorname{Hom}}$ is diagonal and non-zero only in correspondence of the MIs which require a square-root normalisation. 
This enables the straightforward use of finite fields techniques in the reconstruction of DEs~\cite{Peraro:2019svx}. 
\end{itemize}
We perform the previous steps sector by sector, starting from the lower sectors and going up to the top sectors.
Finally, we proceed with the last step.
\begin{itemize}
\item \textbf{Step 3}: We reconstruct analytically the DEs, this time including also the sub-sectors contributions, with respect to the basis $\vec{\mathcal{J}}_{F} = \cup_S \vec{\mathcal{J}}_{F,S}$, once again keeping the square roots out of the computation.  
While for most of the sectors the previous steps
are enough to ensure that also the sub-sectors part of the DEs is in $\epsilon$-factorised form, we found that some integral sectors need further adjustments. In this case, 
in order to achieve an $\epsilon$-factorised form, it is sufficient to modify the definition of the MIs in the sector by including appropriate linear combination of the MIs
of the sub-sectors which are not in $\epsilon$-factorised form. The specific form of the linear combinations is fixed by demanding that the DEs are $\epsilon$-factorised off diagonal (e.g.\ see~\cite{Gehrmann:2014bfa}). 
\end{itemize}
As a result, we obtain DEs whose connection matrices are rational, $\eps$-factorised off diagonal, and linear in $\eps$ on the diagonal.
The $\eps$-factorised form can then be obtained via the rotation
$\vec{\mathcal{I}}_{F} = N_{F}(\vec{x}) \cdot \vec{\mathcal{J}}_{F}$, as
\begin{equation}
  \dd\, \vec{\mathcal{I}}_{F}(\vec{x},\eps) = \eps \left( N_{F}(\vec{x}) \cdot  \dd \widehat{A}^{(F)}_{1,\operatorname{Hom}}(\vec{x}) \cdot N_{F}^{-1}(\vec{x}) \right) \cdot \vec{\mathcal{I}}_{F}(\vec{x},\eps) \,,
\end{equation}
where $N_{F}(\vec{x})$ is a diagonal matrix which captures all square-root normalisations and satisfies
\begin{equation}
\dd \widehat{A}^{(F)}_{0,\operatorname{Hom}} + N_{F}^{-1} \cdot \dd N_{F}  = 0\,.
\end{equation}

Through the strategy described above, we built a basis of MIs for topology ${\rm PB}_C$ which satisfies the DEs in canonical form. 
Contrarily, such a form is not possible for topology ${\rm PB}_B$.
Two sectors, shown in fig.~\ref{fig:probSecB}, present additional challenges.
These can be identified by analysing the factorisation properties of the Picard-Fuchs operators~\cite{Muller-Stach:2012tgj,Adams:2017tga}.
The sectors in fig.~\ref{fig:probSecB} are in fact the only ones whose MIs have Picard-Fuchs operators with irreducible factors of degree $2$, in contrast with all the other MIs, whose Picard-Fuchs operators factorise into linear factors.
We devote sections~\ref{sec:421B} and~\ref{sec:321B} to a thorough analysis of these sectors, and summarise here the main conclusions.
By analysing the homogeneous DEs for the sector in fig.~\ref{fig:321B}, we find that their solution involves elliptic integrals.
While the last few years have seen important progress in the construction of $\eps$-factorised
DEs beyond the standard `$\eps \times \dd \log$' case in eqs.~\eqref{eq:DEsCanonical} and~\eqref{eq:dAdlog}~\cite{Frellesvig:2021hkr,Frellesvig:2023iwr,Gorges:2023zgv}, this problem is challenging in general.
Moreover, the transformation required to achieve an $\eps$-factorised form in this case involves transcendental functions (such as elliptic ones). 
This feature adds a further level of complexity for the numerical evaluation. 
In other words, even if an $\eps$-factorised form of the DEs could be obtained in this case, the numerical evaluation of the solution would remain an open problem.
The most common approach in such cases is to resort to semi-numerical methods such as the generalised power series expansion, which can be equally applied even without an $\eps$-factorised form.
Nonetheless, we put some effort into choosing MIs for this sector such that the connection matrices are polynomial in $\eps$ up to order $\eps^2$, and all the entries which do not involve MIs of the two problematic sectors are $\eps$-factorised.
With respect to the generic form, this makes the expression of the DEs more compact, and improves the evaluation time of the solution using generalised power series expansions.

The second problematic sector, shown in fig.~\ref{fig:421B}, can be put into $\eps$-factorised form, but at the cost of introducing a nested square root, similar to the one encountered in ref.~\cite{FebresCordero:2023gjh}. 
While in principle this is not a problem, in practice the available codes implementing the generalised power series expansion method cannot handle a nested square root. 
For this reason, we prefer to omit this transformation, and adopt a basis of $3$ MIs for this sector such that the connection matrices have a $2\times 2$ block with non-zero $\eps^0$ terms.

In light of the previous considerations, we built a basis of MIs for topology ${\rm PB}_B$ which satisfies a system of DEs as in eq.~\eqref{eq:DEsGeneral} with the connection matrix $\dd A^{({\rm PB}_B)}$ of the form given in eq.~\eqref{eq:DEsPolyEps} with $k_{\rm max} = 2$.
Only the 27 entries of the connection matrix which couple the differential of $\mathcal{I}_{{\rm PB}_B,37}$ to MIs other than itself are quadratic in $\eps$. 
These entries depend on a subset of the kinematic invariants (see sec.~\ref{sec:321B}).
The entries which are instead non-zero at $\eps=0$ are 16. Of these, 4 involve the MIs $\mathcal{I}_{{\rm PB}_B,19}$ and $\mathcal{I}_{{\rm PB}_B,20}$, and can be eliminated as discussed in sec.~\ref{sec:421B} at the cost of introducing a nested square root. 
The remaining entries involve at least one of the MIs of the elliptic sector (see sec.~\ref{sec:321B}).

Before we move on to discussing the choice of MIs for the most complicated sectors, we remark that
another important aspect in the construction of the integral bases is to minimise both the highest numerator rank and the quantity of dotted propagators.
This requirement is crucial in order to prevent the size and number of IBP relations needed from exploding and thus making the calculation computationally too expensive.
To further ameliorate this aspect, we used 
the software \textsc{NeatIBP}~\cite{Wu:2023upw} to generate optimised IBP relations through the solution of syzygy equations~\cite{Gluza:2010ws}.

\subsection{Pentagon-box sectors}

The eight-propagator pentagon-box sectors shown in figures~\ref{fig:PBttjB}
and~\ref{fig:PBttjC} contain three MIs for topology ${\rm PB}_B$ and four MIs
for topology ${\rm PB}_C$. 

Regarding topology ${\rm PB}_B$, since the number of MIs is the same as in the
easier mass configurations, we then find that a canonical basis of MIs for
this sector is~\cite{Badger:2022hno}
\begin{align}
\label{eq:PBB-topsec}
\begin{aligned}
\cI_{{\rm PB}_B,1} &= \epsilon^4 \, d_{15} \, \operatorname{tr}_5 I_{1,1,1,1,1,1,1,1}^{({\rm PB}_B),[12],0,0,0} \,, \\
\cI_{{\rm PB}_B,2} &= \epsilon^4 \, d_{15} \, \operatorname{tr}_5  I_{1,1,1,1,1,1,1,1}^{({\rm PB}_B),[11],0,0,0} \,,\\ 
  \cI_{{\rm PB}_B,3} &= \epsilon^4 \, d_{15} \, d_{23}\,d_{34} \left(I_{1,1,1,1,1,1,1,1}^{({\rm PB}_B),1,0,0} + m_t^2 \, I_{1,1,1,1,1,1,1,1}^{({\rm PB}_B),0,0,0}\right) \,.
\end{aligned}
\end{align}
From numerical evaluations with \textsc{AMFlow} we observe that $\cI_{{\rm
PB}_B,1}$ vanishes up to order $\eps^4$.

Topology ${\rm PB}_C$ instead has four MIs in the top sector.  We choose the
first three similarly to topology ${\rm PB}_B$.  The construction of the fourth MI is
more complicated.  We start from the scalar integral in $d=6-2 \eps$
dimensions, expressed in terms of integrals in $d=4-2\eps$ dimensions using
\textsc{LiteRed}'s implementation of the dimension-shifting
relations~\cite{Tarasov:1996br}.  With this choice, the DEs are linear in
$\eps$.  We then construct a transformation of the fourth MI to eliminate the
$\eps^0$ terms.  The resulting expression is however complicated, has rank-$4$
numerators, and involves a spurious pole.  We then search for a representation
of this MI which is free of these undesirable features by fitting an ansatz
made of (at most) rank-$2$ integrals on the top sector. This resulted in the following choices:
\begin{align}
\begin{aligned}
\cI_{{\rm PB}_C,1} &= \eps^4 \, \, d_{34} \, d_{45} \left(d_{12}+m_t^2\right) I_{1,1,1,1,1,1,1,1}^{({\rm PB}_C),1,0,0} \,,\\ 
\cI_{{\rm PB}_C,2} &= \eps^4 \, \operatorname{tr}_5 \, (d_{12} + m_t^2) \, I_{1,1,1,1,1,1,1,1}^{({\rm PB}_C),[11],0,0,0} \,,\\ 
\cI_{{\rm PB}_C,3} &= \eps^4 \,  \operatorname{tr}_5 \, (d_{12} + m_t^2)  \, I_{1,1,1,1,1,1,1,1}^{({\rm PB}_C),[12],0,0,0} \,,\\
\cI_{{\rm PB}_C,4} &=  \eps^4 \, \frac{\beta_{12}}{\operatorname{tr}_5} \,4 \, d_{34} d_{45} (d_{12} + m_t^2) \bigl[(d_{45} -d_{12} - m_t^2) I_{1,1,1,1,1,1,1,1}^{({\rm PB}_C),1,0,1} - d_{34} d_{45} I_{1,1,1,1,1,1,1,1}^{({\rm PB}_C),1,0,0} + \\ 
& \phantom{{} = {}}  2 d_{23} (d_{12} + m_t^2) I_{1,1,1,1,1,1,1,1}^{({\rm PB}_C),0,0,1}	 + (\text{sub-sectors}) \bigr] \,.
\end{aligned}
\end{align}
The sub-sector terms in $\cI_{{\rm PB}_C,4}$ are rather lengthy, and can be
found in the ancillary files~\cite{ancillary}.  Interestingly, numerical
evaluations with \textsc{AMFlow} show that three of the four MIs of this
sector ($\cI_{{\rm PB}_C,2}$, $\cI_{{\rm PB}_C,3}$ and $\cI_{{\rm PB}_C,4}$)
vanish up to order $\eps^4$.

\subsection{Double-box sectors}

\begin{figure}[h]
	\centering
	\begin{subfigure}{0.22\linewidth}
		\includegraphics[width=\linewidth]{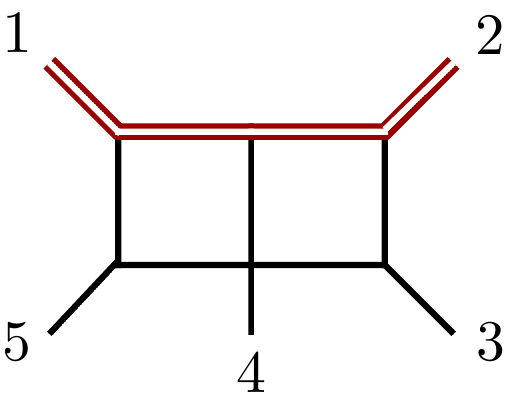}
	 	\caption{}
		\label{fig:331aB}
	\end{subfigure}
	\begin{subfigure}{0.22\linewidth}
		\includegraphics[width=\linewidth]{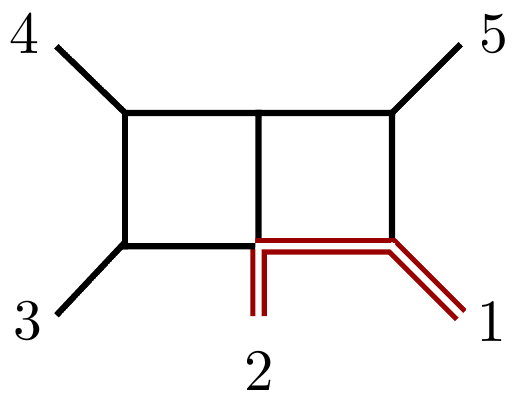}
		\caption{}
		\label{fig:331bB}
	\end{subfigure}
	\begin{subfigure}{0.22\linewidth}
		\includegraphics[width=\linewidth]{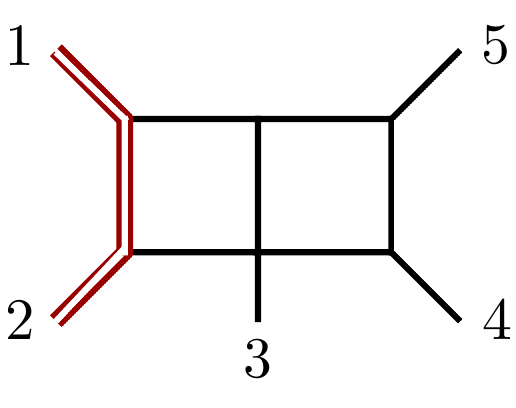}
		\caption{}
		\label{fig:331aC}
	\end{subfigure}
	\begin{subfigure}{0.22\linewidth}
		\includegraphics[width=\linewidth]{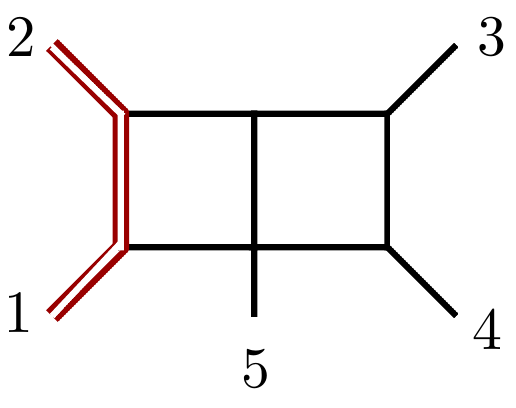}
		\caption{}
		\label{fig:331dC}
	\end{subfigure}
	\caption{The four five-point double-box topologies covering the MIs $\cI_{{\rm PB}_B,4}$ -- $\cI_{{\rm PB}_B,9}$~(a), $\cI_{{\rm PB}_B,10} $ -- $\cI_{{\rm PB}_B,12}$~(b), $\cI_{{\rm PB}_C,5} $ -- $\cI_{{\rm PB}_C,8}$~(c) and $\cI_{{\rm PB}_C,15} $ -- $\cI_{{\rm PB}_C,18}$~(d) respectively.}
	\label{fig:db}
\end{figure}

There are four double-box sectors, two in topology ${\rm PB}_B$ and two in
topology ${\rm PB}_C$, as shown in fig.~\ref{fig:db}.  The sectors~(b), (c)
and~(d) in fig.~\ref{fig:db} have already been discussed in
ref.~\cite{Badger:2022hno}. Sector~(a) of topology ${\rm PB}_B$ is instead
new and contains six MIs. As for the pentagon-box sectors, we were able to construct compact expressions for some
of the canonical MIs of this sector using $\mu_{ij}$ numerator insertions.
The first three integrals ($\cI_{{\rm PB}_B,4}$,
$\cI_{{\rm PB}_B,5}$ and $\cI_{{\rm PB}_B,6}$) can be chosen as in sector~(b)
in fig.~\ref{fig:db} (see ref.~\cite{Badger:2022hno}).  Of the remaining
three, we defined two ($\cI_{{\rm PB}_B,7}$ and $\cI_{{\rm PB}_B,8}$) using
$\mu_{ij}$ numerators and dotted propagators:
\begin{align} 
\begin{aligned}
  \cI_{{\rm PB}_B,4}&= \epsilon^4 \, d_{15} \, d_{23} \, (d_{12} +m_t^2) I_{1,1,1,0,1,1,1,1}^{({\rm PB}_B),0,0,0} \,,\\ 
  \cI_{{\rm PB}_B,5}&= \epsilon^4 \, d_{15} \, d_{23}\, \left(I_{1,1,1,0,1,1,1,1}^{({\rm PB}_B),0,1,0} + m_t^2 \, I_{1,1,1,0,1,1,1,1}^{({\rm PB}_B),0,0,0}\right) \,,\\ 
  \cI_{{\rm PB}_B,6}&= \epsilon^4 \, \operatorname{tr}_5 \, \left( I_{1,1,1,0,1,1,1,1}^{({\rm PB}_B),[12],0,0,0} + (\text{sub-sectors}) \right) \,, \\
  \cI_{{\rm PB}_B,7}&= \epsilon^3 \, d_{23} \, \operatorname{tr}_5 \, \left( I_{1,1,2,0,1,1,1,1}^{({\rm PB}_B),[12],0,0,0} + (\text{sub-sectors}) \right) \,,\\ 
  \cI_{{\rm PB}_B,8}&= \epsilon^3 \, d_{15} \, \operatorname{tr}_5 \, I_{1,1,1,0,1,1,2,1}^{({\rm PB}_B),[12],0,0,0} \,. \\
\end{aligned}
\end{align}
From numerical evaluations with \textsc{AMFlow}, we observe that $\cI_{{\rm PB}_B,6}$ is zero up to order $\eps^4$.
For the sixth MI of this sector ($\cI_{{\rm PB}_B,9}$), we could not find a compact representation. 
We started from the derivative of $\cI_{{\rm PB}_B,4}$ with respect to $d_{23}$, which leads to a linear dependence of the connection matrices on $\eps$. 
We then constructed a transformation to eliminate the $\eps^0$ part of the connection matrices, both in the homogeneous part of the DEs and in the sub-sectors.
This constraint amounts to first-order DEs for the entries of the transformation matrix, which we could solve in terms of rational functions.
The resulting expression for $\cI_{{\rm PB}_B,9}$ is rather lengthy, and can be found in the ancillary files~\cite{ancillary} together with the sub-sector terms of $\cI_{{\rm PB}_B,6}$ and $\cI_{{\rm PB}_B,7}$.

\subsection{Pentagon-triangle sector: a nested square root}
\label{sec:421B}

There is only one pentagon-triangle sector, shown in figure \ref{fig:421B} and
dubbed $421B$, and it appears in topology ${\rm PB}_B$.  This sector has not been studied previously in the literature and
contains three MIs.
One MI can be chosen to take the same form as in the analogous topology of the
five-point integrals with an off-shell leg and massless internal
propagators~\cite{Abreu:2020jxa}. For the remaining two integrals, we made use
of numerator structures inspired by local
numerators~\cite{Arkani-Hamed:2010pyv} and written in terms of Dirac traces, as
\begin{align}
\label{eq:MIs331B}
\begin{aligned}
  \cI_{{\rm PB}_B,18}&= \epsilon^4 \, \operatorname{tr}_5 \, I_{1,1,1,1,1,1,0,1}^{({\rm PB}_B),[11],0,0,0} \,, \\
  \cI_{{\rm PB}_B,19}&= \eps^4 \, d_{45} \, I_{1,1,1,1,1,1,0,1}^{({\rm PB}_B),0,0,0} \left[ {\rm tr}\bigl(\gamma_5 \slashed{p}_3 (\slashed{k}_1- \slashed{p}_2-\slashed{p}_3) \slashed{p}_4 \slashed{p}_2 \bigr)  \right] \,, \\
  \cI_{{\rm PB}_B,20}&= \eps^4 \, d_{45} \, I_{1,1,1,1,1,1,0,1}^{({\rm PB}_B),0,0,0} \left[ {\rm tr}\bigl( \slashed{p}_3 (\slashed{k}_1- \slashed{p}_2-\slashed{p}_3) \slashed{p}_4 \slashed{p}_2 \bigr)  \right] \,,
\end{aligned}
\end{align}  
where the terms in the square brackets are meant to be taken under the integral
sign.  Note that the numerators in $\cI_{{\rm PB}_B,19}$ and $ \cI_{{\rm
PB}_B,20}$ are the parity even and odd parts of the spinor chain $\langle 3 |
k_1 -p_2 - p_3 | 4]$, multiplied by an arbitrary factor of $\langle 4 2 \rangle
[23]$ to cancel the helicity little-group scaling.
From numerical evaluations with \textsc{AMFlow}, we observe that $\cI_{{\rm
PB}_B,19}$ and $\cI_{{\rm PB}_B,20}$ start at order $\eps^4$, whereas
$\cI_{{\rm PB}_B,18}$ vanishes up to order $\eps^4$.

The choice of Dirac trace in the numerator is made in order to cancel the potential
singularity as the propagator momentum $k_1-p_2-p_3$ becomes collinear to
either adjacent massless leg $p_3$ or $p_4$. The two simple numerator
structures that can achieve this are $\spAB{3}{k_1-p_2-p_3}{4}$ and its spinor
conjugate $\spAB{4}{k_1-p_2-p_3}{3}$. Setting these objects inside a trace
ensures they are free from any spinor phases. We choose ${\rm
tr}_\pm{[\slashed{p}_3 (\slashed{k}_1- \slashed{p}_2-\slashed{p}_3)
\slashed{p}_4 \slashed{p}_2]}$, which have the same loop-momentum dependence but
different normalisation. It is then useful to split into parity odd and even
pieces by taking linear combinations. One can also write an alternative version of this `local' numerator as $(l-l^*)^2$, where $l=k_1-p_2-p_3$ and $l^*$
solves the quadruple cut constraints 
\begin{align}
\bigl\{(l^*)^2=0 \,, \ (l^*-p_4)^2=0 \,, \ (l^*+p_3)^2=0 \,, \ (l^*+p_2+p_3)^2=m_t^2 \bigr\} \,.
\end{align}
As discussed in sec.~\ref{sec:setup}, we remove the parity degree of freedom and replace the parity-odd trace with the parity-even square root $\trfive$ defined in eq.~\eqref{eq:tr5}.

With the MIs in eq.~\eqref{eq:MIs331B}, the homogeneous DEs for this sector are linear in $\eps$, but not $\eps$-factorised. 
They take the form
\begin{align} \label{eq:DE331B}
\dd \, \cI_{421B} = \left[ \begin{pmatrix} 0 & 0 & 0 \\ 0 & X & Y \\ 0 & Y & X \end{pmatrix} + \mathcal{O}(\eps) \right] \cdot \cI_{421B} + (\text{sub-sectors}) \,, \qquad \cI_{421B} = \begin{pmatrix} \cI_{{\rm PB}_B,18} \\ \cI_{{\rm PB}_B,19} \\ \cI_{{\rm PB}_B,20} \end{pmatrix} \,.
\end{align}
The omitted terms in the square brackets are proportional to $\eps$. 
All sub-sector terms are $\eps$-factorised, except for the entries coupling $\cI_{{\rm PB}_B,19}$ and $\cI_{{\rm PB}_B,20}$ to the elliptic sector ($\cI_{{\rm PB}_B,i}$ for $i=35,36,37$).
As anticipated in the introduction to this section, the factorisation of $\eps$ in the entire diagonal block requires a transformation involving a nested square root.

The higher complexity of this sector can be detected by analysing the corresponding Picard-Fuchs operators as proposed in ref.~\cite{Adams:2017tga}.
First, we reduce the multi-scale problem to a single-scale one by defining a univariate phase-space slice,
\begin{align} \label{eq:unislice}
x_i = a_i + b_i \, \lambda \,, \qquad \forall \, i=1,\ldots,6\,,
\end{align}
with $a_i, b_i \in \mathbb{Q}$, and viewing the MIs as functions of $\lambda$.
The constants $a_i$ and $b_i$ in eq.~\eqref{eq:unislice} are chosen randomly, but one must make sure that no denominator factors of the connection matrices vanish on the univariate slice, so as to avoid singular points.
Furthermore, we work modulo sub-topologies and modulo $\eps$-corrections, i.e.\ we focus on the $3\times 3$ block of the DEs corresponding to this sector and set $\eps=0$.
The first MI of this sector, $\mathcal{I}_{{\rm PB}_B,18}$, decouples, as the corresponding DE-entries are already $\eps$-factorised. 
In order to decouple $\mathcal{I}_{{\rm PB}_B,19}$ and $\mathcal{I}_{{\rm PB}_B,20}$ we need to differentiate one more time.
In other words, the remaining $2\times 2$ block is equivalent to a second-order ordinary differential equation in $\lambda$ for each of the integrals separately:
\begin{align} \label{eq:PF-19,20}
L_i \, {\rm MaxCut} \left[ \mathcal{I}_{{\rm PB}_B,i} \right]_{\eps=0} = 0 \,, \qquad \quad L_i = \sum_{k=0}^2 c_{i,k}(\lambda) \frac{\dd^k}{\dd \lambda^k} \,,  \qquad \quad \forall \, i=19,20\,.
\end{align}
The differential operators $L_i$ which annihilate the MIs are called Picard-Fuchs operators.
The factorisation properties of the Picard-Fuchs operators encode useful information for the factorisation of $\eps$.
Whenever they factor completely into linear factors, the strategy of ref.~\cite{Adams:2017tga} allows one to construct a transformation which puts the DEs in $\eps$-factorised form.
The Picard-Fuchs operators of $\mathcal{I}_{{\rm PB}_B,19}$ and $\mathcal{I}_{{\rm PB}_B,20}$ are however second-order and irreducible.\footnote{We used the \textsc{Maple} command \texttt{DFactor} to factorise the differential operators~\cite{Hoeij1997FactorizationOD}. 
Note that this algorithm factors into differential operators with rational function coefficients, and may therefore miss factorisations involving algebraic coefficients.}
The appearance of an irreducible factor of order greater than one in the factorisation of a Picard-Fuchs operator is an indication that the Feynman integral
cannot be expressed in terms of MPLs~\cite{Weinzierl:2022eaz}.
In this case, however, we find that the solutions to the Picard-Fuchs equations in eq.~\eqref{eq:PF-19,20} ---~equivalently, the solution to the homogeneous DEs for this block~--- do not contain elliptic integrals, but rather a nested square root.
Indeed, we can actually put the DEs in $\eps$-factorised form with an algebraic change of basis, as we discuss below. 
It may therefore be possible to factor these Picard-Fuchs operators into linear factors by allowing for algebraic functions in the coefficients.
In section~\ref{sec:321B} we will see that also the Picard-Fuchs operators of the integrals of the sector shown in fig.~\ref{fig:321B} contain second-order irreducible factors. 
In that case we will however find elliptic integrals in the solutions.
The Picard-Fuchs operators (modulo $\eps$ corrections and sub-sectors) for all the other MIs of both topology ${\rm PB}_B$ and ${\rm PB}_C$ are instead first-order.

We now proceed to put the homogeneous DEs for this sector in $\eps$-factorised form.
Thanks to the particularly symmetric form of the $2 \times 2$ $\eps^0$ block corresponding to $\cI_{{\rm PB}_B,19}$ and $\cI_{{\rm PB}_B,20}$ (see eq.~\eqref{eq:DE331B}), the off-diagonal non-zero entries can be removed by simply replacing these MIs by their sum and difference, i.e.
\begin{align} \label{eq:sumdiff}
 \cI_{421B} \quad \longrightarrow \quad  \cI'_{421B} = T_1 \cdot  \cI_{421B} \,, \qquad T_1 = \begin{pmatrix} 1 & 0 & 0 \\ 0 & 1 & 1 \\ 0 & 1 & -1 \end{pmatrix} \,.
\end{align}
Note that this transformation mixes integrals of different $\trfive$-charge ($\cI_{{\rm PB}_B,19}$ is even, $\cI_{{\rm PB}_B,20}$ is odd), and gives $\langle 3 | (k_1 -p_2 - p_3 )42|3]$ and its parity conjugate as numerators for $\cI'_{{\rm PB}_B,19}$ and $\cI'_{{\rm PB}_B,20}$.
The resulting homogeneous DEs have non-zero $\eps^0$ terms only on the diagonal, which can thus be removed by proper normalisation.
The required normalisation factors are reciprocal of the solutions to the Picard-Fuchs equations in eq.~\eqref{eq:PF-19,20}, and involve a nested square root:
\begin{align} \label{eq:Npm}
N_{\pm} = \frac{\sqrt{n_{\pm}}}{d_{45} \, r_2} \,,
\end{align}
where
\begin{align}
n_{\pm} = d_{23}^2 \operatorname{tr}_5^2-8 r_2 r_4 \pm 4 d_{23} r_3 \operatorname{tr}_5 \,,
\end{align}
with
\begin{align}
\begin{aligned}
& r_2 =  2 d_{23} (d_{23}+d_{34}-d_{15})+d_{34} m_t^2 \,, \\
& r_3 = d_{23} \left[d_{12} (d_{23}-d_{15}) - d_{23} d_{34} + (d_{34}-d_{15}) d_{45} \right] + \left[d_{23} (d_{15} - d_{23} - 2 d_{34}) + \
d_{34} d_{45}\right] m_t^2 - d_{34} m_t^4 \,, \\
& r_4 = 2 d_{12} d_{15} d_{23} d_{45} + \left[ 2 d_{12} d_{23} (d_{23}-d_{15}) - d_{34} (2 d_{23}^2 - 2 d_{23} d_{45} + d_{45}^2) \right] m_t^2 + \\
& \ \phantom{r_4 = } 2 (d_{23}^2 + d_{34} d_{45} -d_{15} d_{23} ) m_t^4 \,.
\end{aligned}
\end{align}
We emphasise that $n_{\pm}$ involves the square root $\operatorname{tr}_5$, and
that $n_{+}$ and $n_{-}$ are related by swapping the sign of
$\operatorname{tr}_5$.  While the expression of the normalisation factors in
eq.~\eqref{eq:Npm} is fairly intricate, it is straightforward to verify that
they do not introduce spurious singularities.  In other words, the product
$n_{+} n_{-}$ factorises in terms of the same factors present in the
denominators of the connection matrices prior to the transformation.
Furthermore, we note that the factorisation of expressions involving square
roots is not unique, and therefore a simpler representation of these
normalisation factors may exist.
In particular, we find that this nested square root can also be written
compactly in terms of traces of gamma matrices,
\begin{align}
\begin{aligned}
  N_\pm = \sqrt{8 \left(
  \frac{ {\rm tr}_\mp(\slashed{p}_3\slashed{p}_2\slashed{p}_{12}\slashed{p}_5)}{{\rm tr}_\mp(\slashed{p}_4\slashed{p}_5\slashed{p}_3\slashed{p}_2)}  + \frac{4m_t^2}{s_{45}}\right)
  \frac{ {\rm tr}_\mp(\slashed{p}_3\slashed{p}_2\slashed{p}_{12}\slashed{p}_5)}{{\rm tr}_\mp(\slashed{p}_4\slashed{p}_5\slashed{p}_3\slashed{p}_2)}}\,,
  \label{eq:nested_tr}
\end{aligned}
\end{align}
in which it is clear that the interior square root could be rationalised when
using a momentum-twistor representation~\cite{Hodges:2009hk,Badger:2022mrb}
with rational parameterisations for the spinor products, and that the argument of the 
outer square root becomes a perfect square in the massless limit. We have checked
that $n_{\pm}$ cannot be expressed as a perfect square of the form $(a+b
\operatorname{tr}_5)^2$, for some rational functions $a$ and $b$, which would
allow us to remove the exterior square root.  Finally, the normalisation
factors in eq.~\eqref{eq:Npm} do not have well-defined behaviour under swapping
the sign of $\operatorname{tr}_5$.  It is instead desirable that all MIs are
either even or odd with respect to this operation.  We therefore apply another
transformation of the form of eq.~\eqref{eq:sumdiff} to restore this property.
In conclusion, the basis of this sector which puts the DEs in $\eps$-factorised
form~is
\begin{align}
\cI''_{421B} = \begin{pmatrix} 1 & 0 & 0 \\ 0 & 1 & 1 \\ 0 & 1 & -1 \end{pmatrix} \cdot 
  \begin{pmatrix} 1 & 0 & 0 \\ 0 & N_+ & 0 \\ 0 & 0 & N_- \end{pmatrix} \cdot 
  \begin{pmatrix} 1 & 0 & 0 \\ 0 & 1 & 1 \\ 0 & 1 & -1 \end{pmatrix} \cdot  \cI_{421B}  \,.
\end{align}
Keeping in mind that we did not put in $\eps$-factorised form the elliptic sector (see sec.~\ref{sec:321B}), and that \textsc{DiffExp} cannot handle nested square roots, we prefer to omit this transformation from our chosen basis of MIs.

\subsection{Pentagon-bubble sectors}

\begin{figure}[H]
	\centering
	\begin{subfigure}{0.3\linewidth}
		\includegraphics[width=\columnwidth]{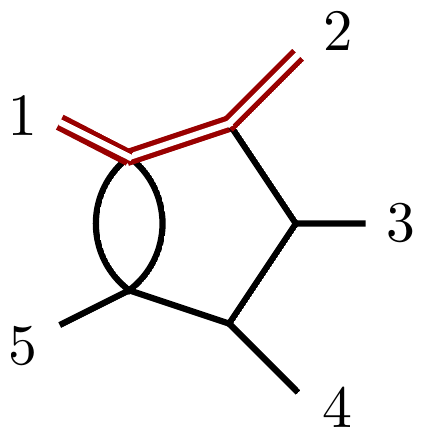}
		\caption{}
		\label{fig:411B}
	\end{subfigure}
	\hspace{3cm}
	\begin{subfigure}{0.3\linewidth}
		\includegraphics[width=\columnwidth]{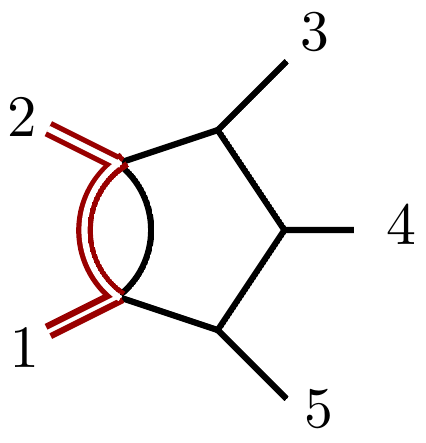}
		\caption{}
		\label{fig:411C}
	\end{subfigure}
         \caption{The two pentagon-bubble sectors for topologies ${\rm PB}_B$ (sub-figure~(a)) and ${\rm PB}_C$ (sub-figure~(b)).}
	\label{fig:pentagon-bubble}
\end{figure}

There are two sectors in the form of a pentagon with a bubble insertion, shown in figure~\ref{fig:pentagon-bubble}: one in topology ${\rm PB}_B$, with two MIs, and one in topology ${\rm PB}_C$, with three MIs.
\noindent
Both sectors present a new mass configuration.
However, since the number of MIs for the pentagon-bubble integrals in topology ${\rm PB}_B$ is the same as in previously studied cases, we can make a similar choice for the canonical basis:
\begin{align}
\begin{aligned}
\cI_{{\rm PB}_B,45}&= \epsilon^3\, \operatorname{tr}_5 \, I_{1,1,1,1,0,1,0,2}^{({\rm PB}_B),[11],0,0,0} \,,\\  
\cI_{{\rm PB}_B,46}&= \epsilon^3(1 - 2\epsilon)\, d_{23}\, d_{34}  I_{1,1,1,1,0,1,0,1}^{({\rm PB}_B),0,0,0} \,.
\end{aligned}
\end{align}
Regarding the pentagon-bubble sector in topology ${\rm PB}_C$, we find a canonical basis where all three MIs involve dotted propagators:
\begin{align}
\begin{aligned}
\cI_{{\rm PB}_C,32}&= \epsilon^3\, \operatorname{tr}_5 \, I_{1,1,1,1,0,2,0,1}^{({\rm PB}_C),[11],0,0,0} \,,\\   
\cI_{{\rm PB}_C,33}&= \epsilon^3\, \operatorname{tr}_5 \, I_{1,1,1,1,0,1,0,2}^{({\rm PB}_C),[11],0,0,0} \,,\\   
\cI_{{\rm PB}_C,34}&= \epsilon^3 \,d_{34}\, d_{45}  \left( I_{1,1,1,1,0,2,0,1}^{({\rm PB}_C),1,0,0} + m_t^2 I_{1,1,1,1,0,2,0,1}^{({\rm PB}_C),0,0,0}\right) \,.
\end{aligned}
\end{align}

\subsection{Elliptic sector}
\label{sec:321B}

The most complicated sector belongs to topology ${\rm PB}_B$ and is shown in fig.~\ref{fig:321B}. 
It is a four-point sector, and thus its integrals depends on four variables only ($d_{12}$, $d_{34}$, $d_{15}$ and $m_t^2$).
There are $3$ MIs, which we choose as
\begin{align}
\label{eq:321B}
\begin{aligned}
\cI_{{\rm PB}_B,35}&= \eps^4 \, d_{15} \, (d_{12} + m_t^2) \, I_{1,1,0,1,1,1,0,1}^{({\rm PB}_B),0,0,0} \,,\\  
\cI_{{\rm PB}_B,36}&= \eps^4 \, \sqrt{(d_{15}-d_{34})^2-2 d_{34} m_t^2} \, I_{1,1,0,1,1,1,0,1}^{({\rm PB}_B),0,0,0} \left[2 k_1 \cdot p_1 \right] \,, \\
\cI_{{\rm PB}_B,37}&= \eps^4 \, (-1+2\eps) \, d_{15} \, I_{1,1,0,1,1,1,0,1}^{({\rm PB}_B),0,0,0} \left[2 k_2 \cdot p_2 \right] \,,
\end{aligned}
\end{align}
where we recall that the terms in the square brackets are meant to be taken under the integral sign.
From numerical evaluations with \textsc{AMFlow}, we observe that the chosen MIs of this sector are non-zero only starting from order $\eps^4$.
While the DEs for this sector are not $\eps$-factorised, the MIs above simplify them substantially with respect to an arbitrary choice.
The normalisation factor of $\cI_{{\rm PB}_B,36}$ is chosen so as to factorise $\eps$ in the corresponding diagonal entry of the DEs.
The factor of $(-1+2\eps)$ in $\cI_{{\rm PB}_B,37}$ is inserted to remove all $\eps$-dependent factors from the denominators of the connection matrices.
The remaining kinematic-dependent normalisation factors of $\cI_{{\rm PB}_B,35}$ and $\cI_{{\rm PB}_B,37}$ ensure that all MIs in the basis have the same dimensionality, and are chosen heuristically as they lead to more compact connection matrices.
The DEs have the following structure:
\begin{align}
\dd \, \mathcal{I}_{321B} = \begin{pmatrix} * + \eps \, * & \eps \, *	& * \\
 *+ \eps \, * & \eps \, *	& * \\ 
 (1-2 \eps) (*+\eps \, *) \qquad & (1-2 \eps) \eps \, * \qquad &	*+\eps \, * \\  \end{pmatrix} \cdot \, \mathcal{I}_{321B} + (\text{sub-sectors}) \,,
\end{align}
where $\mathcal{I}_{321B} = \left(\cI_{{\rm PB}_B,35}, \, \cI_{{\rm PB}_B,36}, \, \cI_{{\rm PB}_B,37}\right)^{\top} $, and each asterisk denotes a distinct one-form.
The sub-sectors follow the same pattern as the diagonal block shown above: the entries coupling $\cI_{{\rm PB}_B,35}$ and $\cI_{{\rm PB}_B,36}$ to the sub-sectors are linear in $\eps$,  while for $\cI_{{\rm PB}_B,37}$ they are quadratic.

The analysis of the Picard-Fuchs operators $L_i$ allows us to better characterise the complexity of this sector.
Following the procedure outlined in sec.~\ref{sec:421B}, we construct the differential operators $L_i$ such~that
\begin{align} \label{eq:PF-19,20}
L_i \, {\rm MaxCut} \left[ \mathcal{I}_{{\rm PB}_B,i} \right]_{\eps=0} = 0 \,, \qquad \quad L_i = \sum_{k=0}^{r_i} c_{i,k}(\lambda) \frac{\dd^k}{\dd \lambda^k} \,,  \qquad \quad \forall \, i=35,36,37 \,,
\end{align}
with $r_{35}=r_{37}=2$ and $r_{36}=3$.
We recall that we work on a random univariate phase-space slice (see eq.~\eqref{eq:unislice}), modulo sub-sectors and $\eps$-corrections.
$L_{35}$ and $L_{37}$ are second-order, irreducible operators.
The solution to $L_{35}$ involves the elliptic integral of the first kind ${\rm K}(x)$, while the solution to $L_{37}$ also contains derivatives of the latter.\footnote{We obtained the solutions to the second-order Picard-Fuchs operators with the \textsc{Maple} command \texttt{hypergeometricsols}. We thank Christoph Dlapa for suggesting this.}
$L_{36}$ is instead third-order, and factorises into the product of a second- and a first-order operator.
The first-order factor is simply $\dd/\dd \lambda$, which follows from the fact that the reciprocal of the chosen normalisation factor of $\mathcal{I}_{{\rm PB}_B,36}$ is a solution to $L_{36}$.
In appendix~\ref{sec:CutBaikov} we determine the elliptic curve underlying the elliptic integrals appearing in this analysis, and identify the solutions to the Picard-Fuchs operator $L_{35}$ with the periods of said elliptic~curve.

\section{`$\dd \log$' and `one-form' representation of the differential equations} \label{sec:oneforms}

Having obtained the connection matrices of the differential equations using the integral bases described in the previous sections, we present
compact analytic expressions in terms of independent `$\dd \log$' and `one-form' structures, as in eqs.~\eqref{eq:dAdlog}
and~\eqref{eq:DEsPolyEps}. There are a variety of
methods proposed in the literature to determine the alphabet of `$\dd \log$' forms where they exist. In our case, rather than constructing an ansatz
of possible alphabet letters \textit{a priori}, we first established a set of linearly independent one-forms organised according to
the square root charges defined earlier in section~\ref{sec:deqs}. As discussed there, each entry of the connection matrices after the
reconstruction is a rational function of the invariants, $\vec{x}$. We then add back the square root normalisations of each master integral to rotate
the DEs into $\eps$-factorised form (or as far as possible, in the case of topology ${\rm PB}_B$). After this stage each entry takes the form 
\begin{align}
  \begin{aligned}
  \dd A^{(F)}_{ij}(\vec{x}, \eps) &= \sum_{a} \eps^a \dd A^{(F),a}_{ij}(\vec{x}) \,,\\
  \dd A^{(F),a}_{ij}(\vec{x}) &= \mathcal{C}^{(F)}_{ij}(\vec{x}) \sum_{k=1}^6 \dd x_k \, f^{(F),a}_{ijk}(\vec{x}) \,,
  \end{aligned}
\end{align}
where $\mathcal{C}_{ij}^{(F)}(\vec{x})$ are monomials in the possible square roots (including $\mathcal{C}_{ij}^{(F)}(\vec{x})=1$) that we have called charges (see sec.~\ref{sec:deqs}), and $f^{(F),a}_{ijk}$ are rational
functions of the invariants $\vec{x}$.
The sum over the orders in $\eps$ (index $a$) runs from $0$ to $2$ for ${\rm PB}_B$, while ${\rm PB}_A$ and ${\rm PB}_C$ are in $\eps$-factorised form (i.e., $a=1$). There are 49 charges in total, with 23 appearing in the leading colour topologies. 

At this stage, we determine the linear relations amongst each set of entries $\dd A^{(F),a}_{ij}$ and of $\dd \log W$'s that share the same charge $\mathcal{C}^{(F)}_{ij}$, where we
use the notation $W$ to indicate the letters of our alphabet. Since the square roots are an overall factor, we can divide them out and use finite field
techniques to determine the linear relations. We then solve the linear relations by ordering with respect to the polynomial degree of the entries $\dd A^{(F),a}_{ij}$ and preferring $\dd
\log$ structures. After iterating over all possible charges, all entries $\dd A^{(F),a}_{ij}$ are expressed in terms $\dd \log$ structures of our
alphabet if possible (as in eq.~\eqref{eq:dAdlog}), and additionally a minimal set of simple one-forms otherwise (as in eq.~\eqref{eq:DEsPolyEps}, with some $\omega_i$'s being $\dd \log$'s). 
For concreteness, we spell out the form of the connection matrix for ${\rm PB}_B$:
\begin{align}
\dd A^{({\rm PB}_B)}(\vec{x}, \eps) = \sum_{k=0}^{2} \eps^k \Biggl[ \sum_i  \, c^{({\rm PB}_B)}_{k,i} \, \dd \log\left(W_i(\vec{x})\right) +
   \sum_j \, d^{({\rm PB}_B)}_{k,j} \, \omega_j(\vec{x}) \Biggr] \,,
\end{align}
where $c^{({\rm PB}_B)}_{k,i}$ and $d^{({\rm PB}_B)}_{k,j}$ are matrices of rational numbers.
We stress that the $\dd \log$'s are one-forms themselves, but in this context we call one-forms only those we could not express in terms of $\dd \log$'s.
Furthermore, we emphasise that some of these one-forms are not closed, which implies they are not exact either.\footnote{A differential form $\omega$ is closed if $\dd \omega = 0$. A one-form $\omega$ is exact if there exists a function $f$ such that $\omega = \dd f$. An exact form is thus by definition closed.}
In other words, they cannot be expressed as the differential of any function, let alone of a logarithm.
Table~\ref{tab:dlogandofsummary} shows a summary of each topology.

The rational letters (i.e.\ those with $\mathcal{C}^{(F)}_{ij}=1$) are easy to determine: they are the (algebraically independent) factors in the denominators
of the connection matrices. Some of these letters are extremely simple, linear combinations of invariants, while others are high degree polynomials in the invariants $\vec{x}$ (up to degree $5$). We determine that a large number of these high degree polynomials can be identified as Gram and Cayley determinants of the external
kinematics. Others can be written in compact notation by using traces of gamma matrices. In total we find 58 rational letters for all one-loop and
planar two-loop cases, although only 46 appear in the leading colour topologies.

The algebraic letters are all written in the manifestly odd form
\begin{equation}
  \frac{A+\mathcal{C}}{A-\mathcal{C}} \,,
\end{equation}
where $\mathcal{C}$ is one of the charges (apart from $1$) and $A$ is a rational
function. There are a variety of methods proposed to determine the form of such
letters, for
example~\cite{Zoia:2021zmb,Heller:2019gkq,FebresCordero:2023gjh,Jiang:2024eaj,Henn:2024ngj}.
In our case, we were able to find all the required expressions by comparison
with ans\"atze for the rational function $A$. Rather than printing the list of
charges and letters here we present all the relevant definitions in the
ancillary files~\cite{ancillary} described in appendix~\ref{sec:anc}. In the
case of $\mathcal{C}=\trfive$, we find that compact representations can be found
by using ratios of ${\trp}/{\trm}$ with arguments of either four or six gamma
matrices. Explicit forms are given in the ancillary files, although this is
only an aesthetic consideration.

In some cases, having found $\dd \log$ representations, we have subsequently imposed symmetries amongst the letters. This is particularly relevant for
letters related to charges involving the three-mass triangle Gram determinants, $\Delta_{3,i}$. We note that the charge $\Delta_{3,4}$ appears in
topologies ${\rm PB}_A$ and ${\rm PB}_C$ with 3 independent one-forms. We write the one-forms in terms of 4 letters in order to preserve these
symmetries. This would be convenient in case one was looking for a set of special functions closed under permutations. In addition, for ${\rm PB}_A$ there are two additional letters containing $\trfive$ in our alphabet than independent entries of the connection matrix.

As a final remark, we notice that the mixed $\dd \log$ and one-form expression of the DEs for topology ${\rm PB}_B$ allows us to clearly separate the features in common with the standard canonical cases (${\rm PB}_A$ and ${\rm PB}_C$) from the new, more complicated ones.
Moreover, it makes the expression of the DEs roughly $50$ times more compact than if one stored the connection matrices separately for each derivative, as in the form generated by \textsc{DiffExp} for the semi-numerical solution. It would be interesting to find a way to exploit this simplification of the form of the DEs in the method of generalised power series expansions.

\begin{table}
  \centering
  \begin{tabular}[h]{c|ccc}
  \hline
  topology & charges & $\dd \log$ letters & one-forms  \\
  \hline
  ${\rm PB}_A$ & 17 & 74 & 0 \\
  ${\rm PB}_B$ & 16 & 72 & 63 \\
  ${\rm PB}_C$ & 21 & 80 & 0 \\
  \hline
  2-loop LC & 23 & 98 & 63 \\  
  \hline
  ${\rm P}_A$ & 8 & 41 & 0 \\
  ${\rm P}_B$ & 10 & 45 & 0 \\
  ${\rm P}_C$ & 17 & 55 & 0 \\
  ${\rm P}_D$ & 29 & 64 & 0 \\
  \hline
  1-loop & 40 & 105 & 0 \\
  \hline
  \end{tabular}
  \caption{Summary of the `$\dd \log$' and `one-form' structures appearing in the differential equations for each topology, and cumulatively for all two-loop leading colour (LC) topologies and all 1-loop topologies. ${\rm PB}_A$ and ${\rm PB}_C$ contain three and one more letter
  than linearly independent entries of the connection matrices respectively. This is due to the imposed symmetries described in the text.}
  \label{tab:dlogandofsummary}
\end{table}

\section{Numerical evaluation using generalised series expansions} \label{sec:numerics}

In this section, we discuss the numerical evaluation of the master integrals. 
We obtain a semi-analytic solution to the system of differential equations associated with the master integrals by means of the method of generalised power series expansions~\cite{Francesco:2019yqt}, by exploiting the
\textsc{Mathematica} package \textsc{DiffExp}~\cite{Hidding:2020ytt}. 
We obtain the required boundary values numerically with the package \textsc{AMFlow}~\cite{Liu:2022chg}, which implements the auxiliary mass flow method~\cite{Liu:2017jxz,Liu:2021wks,Liu:2022tji}.
We aim to evaluate the integrals in the physical scattering region relevant for phenomenology.
We begin by defining this region. 
We then motivate our choice for the boundary point, and discuss a number of interesting features of the boundary values.
Next, we present a number of checks we performed to validate our results. 
Finally, we comment on the performance of the numerical evaluation.

\subsection{Physical scattering region}
\label{sec:s45channel}

We restrict our analysis to the physical phase-space region corresponding to the $s_{45}$ scattering channel ($45 \rightarrow 123$).
All other $2\to 3$ channels relevant for $t\bar{t}j$ production can be obtained from the $s_{45}$ channel through suitable permutations of the momenta, and our numerical evaluation procedure can therefore be straightforwardly generalised to them as well.
The $s_{45}$ channel is defined by the following linear constraints on the kinematic invariants,
\begin{align}
\begin{gathered}
p_1^2 > 0 \,, \quad p_2^2 > 0 \,, \quad p_4 \cdot p_5 > 0 \,, \quad p_1 \cdot p_2 > 0 \,, \quad p_1 \cdot p_3 > 0 \,, \quad p_2 \cdot p_3 >0 \,, \\
p_4 \cdot p_1 < 0 \,, \quad p_4 \cdot p_2 < 0 \,, \quad p_4 \cdot p_3 < 0 \,, \quad p_5 \cdot p_1 < 0 \,, \quad p_5 \cdot p_2 < 0 \,, \quad p_5 \cdot p_3 < 0 \,,
\end{gathered}
\end{align}
complemented by the following higher-order constraints coming from Gram determinants,
\begin{align}
\mathrm{det} \, G(p_i, p_j) < 0 \,, \qquad \mathrm{det} \, G(p_i, p_j, p_k) > 0 \,, \qquad  \mathrm{det} \, G(p_i, p_j, p_k, p_l) < 0  \,,
\end{align}
where $i,j,k,l$ take distinct values in $\{1,\ldots,5\}$.
The Gram determinants involving two and four momenta give only one constraint each:
\begin{align}
(d_{12} - m_t^2) (d_{12} + m_t^2) > 0 \,, \qquad \qquad \operatorname{tr}_5^2 < 0 \,.
\end{align}
Those involving three momenta give a number of polynomial constraints on the kinematic invariants which we do not spell out.
All constraints defining the $s_{45}$ channel can be found in ancillary files~\cite{ancillary}.

\subsection{Boundary values}
The evaluation of the master integrals through the method of generalised power series expansions entails the integration of the DEs along a path connecting the target point with a starting point at which the values are known.
We wish to confine such paths to the $s_{45}$ channel in order to avoid analytic continuation, which is non-trivial to determine for multi-variable problems and increases the evaluation time. 
In this view, we choose a boundary point $\vec{x}_0$ in the $s_{45}$ channel.
The choice is arbitrary, but has an impact on the performance of the evaluation.
We choose
\begin{equation} \label{eq:bound_point}
\vec{x}_0 = \bigl\{2,1,-1, 5,-2,1\bigr\} \,,
\end{equation}
following the criteria set in ref.~\cite{Chicherin:2021dyp}, which we recall here.
\begin{enumerate}
\item The point $\vec{x}_0$ is invariant under the symmetries of the $s_{45}$ channel, i.e.\ the exchanges of the external momenta
$p_1 \leftrightarrow p_2$ and $p_4 \leftrightarrow p_5$.
\item The point $\vec{x}_0$ introduces a minimal number of distinct prime factors.
\item The point $\vec{x}_0$ lies on the spurious singularity $d_{23} + d_{34} = 0$. 
\end{enumerate}
The first two criteria reduce the number of independent transcendental constants in the values of the master integrals at $\vec{x}_0$.
This is useful in view of a future analytic solution of the DEs.
The second criterion also makes the numerical evaluation using \textsc{AMFlow} faster, as it reduces the number of prime fields required to reconstruct the DEs with respect to the auxiliary mass.
We recall in fact that we interface \textsc{AMFlow} to \textsc{FiniteFlow} in order to solve the IBP relations over finite fields.

In order to understand the third criterion, we first need to define spurious singularities. 
A \emph{spurious singularity} is a singularity of the connection matrix of the DEs which is not a singularity of the solution to the DEs once the boundary values are taken into account.
The possible singularities of the MIs correspond to the factors in the denominators of the connection matrices.
For DEs where a canonical `$\dd \log$ form' is possible, this is equivalent to the rational letters of the alphabet.
We find that two of them, $d_{23} + d_{34}$ and $d_{12}+d_{15}+m_t^2$, can vanish within the $s_{45}$ channel.\footnote{We can prove analytically that most of the denominator factors have fixed sign in the $s_{45}$ channel. For a few, we have statistical evidence based on $100$K random phase-space points generated by sampling uniformly a parameterisation of the momenta in terms of energies and angles.}
By choosing the boundary point such that $d_{23} + d_{34} = 0$, we ensure that this spurious singularity is never crossed by a straight path starting from $\vec{x}_0$.
This improves the speed of the integration with the generalised power series expansion methods, and is useful in view of a future solution in terms of one-fold integral representations along the lines of ref.~\cite{Caron-Huot:2014lda}.
We emphasise however that the presence of spurious singularities is not an obstacle for the generalised power series expansion method.

We observe that, for a small fraction of phase-space points in the $s_{45}$ channel, the straight line connecting them to $\vec{x}_0$ leaves the $s_{45}$ channel.
A similar structure of the physical phase space is discussed in ref.~\cite{Chicherin:2021dyp}.
In such a case, one can either perform the analytic continuation, or choose a different starting point. 

We use \textsc{AMFlow} to obtain the values of all MIs at $\vec{x}_0$ with, at least, $32$-digit precision.
This gives an upper bound on the achievable precision in the numerical evaluation of the MIs using the results provided in this work.
We expect this level of precision to be sufficient for the application in phenomenology at NNLO in QCD, based on the available experience with two-loop scattering amplitudes with kinematics of similar complexity.
However, higher precision can be easily achieved, if needed, as evaluating the MIs in a single point with \textsc{AMFlow} does not represent a bottleneck for this computation.
We provide the boundary values in the supplementary material~\cite{ancillary}.

For the topologies ${\rm PB}_A$ and ${\rm PB}_C$, the canonical form of the DEs implies that the values at order $\eps^0$ are rational, and can be determined up to the overall normalisation by imposing `first-entry' conditions~\cite{Gaiotto:2011dt}.
In other words, only logarithms of the following arguments can appear at order~$\eps^1$,
\begin{align} \label{eq:allowed1stentries}
\bigl\{m_t^2 \,, \ 2 (d_{12} + m_t^2)\,, \ 2 d_{23} \,, \ 2 d_{34} \,, \ 2 d_{45} \,, \ 2 d_{15} \bigr\}\,,
\end{align}
and this gives linear constraints on the $\mathcal{O}(\eps^0)$ boundary values.
The allowed logarithm arguments can be read off from the graph polynomial $\mathcal{F}$.
We then fix the overall normalisation by rationalising the values from \textsc{AMFlow}.
Similarly, at order $\mathcal{O}(\eps^1)$ we verify using the PSLQ algorithm~\cite{PSLQ} that the values are $\mathbb{Q}$-linear combinations of logarithms of the functions in eq.~\eqref{eq:allowed1stentries} evaluated at $\vec{x}_0$.

As for topology ${\rm PB}_B$, although the DEs are not in canonical form, we still observe that the boundary values are rational at order $\eps^0$, and linear combinations of the logarithms above at order $\eps^1$. 
This follows from the fact that the `problematic' MIs, namely those of the sector involving the nested square root ($\mathcal{I}_{{\rm PB}_B,19}$ and $\mathcal{I}_{{\rm PB}_B,20}$, see sec.~\ref{sec:421B}) and those of the sector involving elliptic integrals ($\mathcal{I}_{{\rm PB}_B,35}$, $\mathcal{I}_{{\rm PB}_B,36}$ and $\mathcal{I}_{{\rm PB}_B,37}$, see sec.~\ref{sec:321B}), are non-zero only starting from order~$\eps^4$.

\subsection{Checks}

The rationality of the boundary values at order $\eps^0$ and the first-entry conditions discussed in the previous subsection are already non-trivial checks of our results.
In order to validate more robustly the numerical evaluation of the MIs, we compared the numerical values obtained by integrating the DEs with \textsc{DiffExp} starting from $\vec{x}_0$ against numerical
evaluations performed with \textsc{AMFlow} at a number of points in the $s_{45}$ channel. 
We found full agreement within the accuracy estimated by \textsc{DiffExp} and \textsc{AMFlow}.
In particular, we perform this check at the following random point,
\begin{equation} \label{eq:x1}
\vec{x}_1 = \left\{\frac{4602}{57095}, \frac{217}{8151}, -\frac{8513}{67193}, \frac{7}{22}, -\frac{14291}{77626}, \frac{1701}{90164}\right\} \,,
\end{equation}
which lies in the $s_{45}$ channel on the other side of the spurious singularity $d_{12} + d_{15} + m_t^2 = 0$ with respect to the boundary point $\vec{x_0}$.
This allows us to check the stability of the numerical evaluation when integrating the DEs over a path that crosses this spurious
singularity.
We provide the values of the master integrals at $\vec{x}_1$ as benchmarks in ancillary files~\cite{ancillary}.

\subsection{Performance analysis}

We finish this section by making some considerations regarding the performance of the numerical evaluation of the MIs. 
First of all, we want to clarify that our evaluation strategy is not optimised for phenomenological applications, and we therefore refrain from making absolute statements about the evaluation time.
The latter in fact depends strongly on the segmentation of the path within the generalised power series expansion method~\cite{Francesco:2019yqt,Hidding:2020ytt}.
The number of segments, in turn, depends on the chosen endpoints of the path, and on the location of the nearest singularities.
An evaluation strategy aimed at a large number of points should therefore minimise the number of segments in the evaluations by re-using iteratively the values obtained with previous evaluations (see e.g.\ ref.~\cite{Abreu:2020jxa}).
We leave this to future work.

\begin{figure}[H]
	\centering
		\includegraphics[width=0.6\columnwidth]{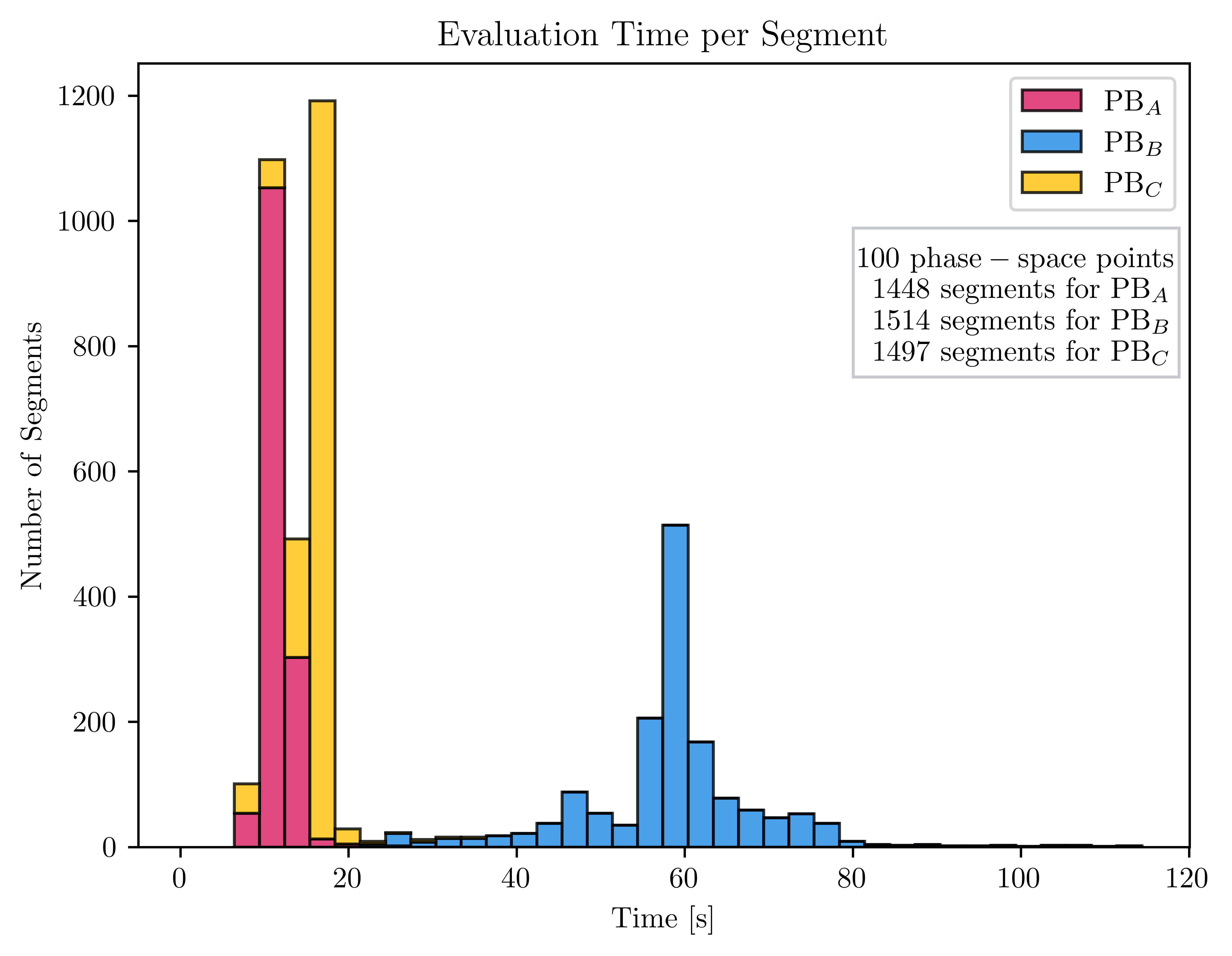}
         \caption{Histogram showing the distribution of the evaluation time per segment for topologies ${\rm PB}_A$, ${\rm PB}_B$ and ${\rm PB}_C$.
         Note that the number of segments depends on the singularity structure and is therefore different for each topology.}
	\label{fig:eval_time}
\end{figure}

Nonetheless, it is still interesting to compare the relative performance of the evaluation of the MIs of topologies ${\rm PB}_A$, ${\rm PB}_C$ and ${\rm PB}_B$, as  the $\epsilon$ structure of the corresponding DEs is different.
We recall that for topology ${\rm PB}_A$ and ${\rm PB}_C$ we have canonical DEs where $\eps$ is factorised, whereas the entries of the connection matrix of topology ${\rm PB}_B$ are degree-2 polynomials in $\eps$.
We can estimate the impact of the additional terms in $\epsilon$ for topology ${\rm PB}_B$ by comparing the evaluation time per segment between topologies for the same values of all parameters.
We performed this analysis on a sample of $100$ phase-space points in the $s_{45}$ channel, amounting to $\approx 1.5$ K segments starting from the boundary point $\vec{x}_0$, with a target accuracy of $10^{-16}$ in $\textsc{DiffExp}$.
The results of this analysis are shown in figure~\ref{fig:eval_time}.\footnote{All the evaluations are performed on an Intel(R) Xeon(R) Gold 5218 2.30~GHz CPU.} 
The different $\epsilon$ structure for the DEs, together with the larger number of MIs, results in a higher evaluation time of ${\rm PB}_B$ with respect to ${\rm PB}_A$ and ${\rm PB}_C$. Indeed, the average evaluation time per segment is $\approx 12$ seconds for ${\rm PB}_A$, $\approx 16$ seconds for ${\rm PB}_C$, and $\approx 58$ seconds for~${\rm PB}_B$.

We finish this section with some comments. First, the data shown in figure~\ref{fig:eval_time} have to be intended as a qualitative analysis and not as a serious attempt of performing 
a statistical study of the evaluation time performance of the method. Furthermore, if we take into account the earlier analysis, bearing in mind that our evaluation strategy is not tailored 
for phenomenological applications, we find the current results to be promising. Finally, it would be interesting to compare the evaluation time of our solution with a possible generalised power series implementation for the case where the connection matrix contains elliptic kernels. Indeed, assuming to be able to build an integral basis in which the DEs are factorised also in the elliptic case, the resulting connection matrix will then depend on elliptic functions. Therefore, it will be interesting to weigh whether is the polynomial dependence on $\epsilon$, or the presence of elliptic functions, to have the most significant impact on the
numerical evaluation of the MIs.

\section{Conclusion} \label{sec:conclusion}

In this work, we have presented compact differential equations for the master integrals of all integral topologies required to describe the production
of a pair of top quarks in association with a jet at hadron colliders at NNLO in leading colour QCD. There were two new pentagon-box topologies to
consider, ${\rm PB}_B$ (fig.~\ref{fig:PBttjB}) and ${\rm PB}_C$ (fig.~\ref{fig:PBttjC}). The latter followed a pattern similar to the one observed in ref.~\cite{Badger:2022mrb} for
${\rm PB}_A$ (fig.~\ref{fig:PBttjA}) and we obtain the DEs in the canonical form, i.e., such that the connection
matrices are given by an overall factor of $\eps$ and $\mathbb{Q}$-linear combinations of logarithmic one-forms ($\dd \log$'s). 
We showed that the former, ${\rm PB}_B$, has new features that prevent it from having the same canonical form. While the one-loop topologies were
previously studied and presented in $\eps$-factorised form in ref.~\cite{Badger:2022mrb}, we present them here for completeness in terms of $\dd
\log$'s.

Topology ${\rm PB}_B$ displays two new features. First, the sector shown graphically in fig.~\ref{fig:421B} involves a nested square root.  Secondly,
the sector in fig.~\ref{fig:321B} involves elliptic integrals, which we identify as the periods of the elliptic curve in
eq.~\eqref{eq:elliptic_curve}. Obtaining canonical DEs in the presence of elliptic integrals is at the forefront of current research and the very
notion of `canonical' in such cases is still under debate. From a practical view point, even if such a generalised canonical form is achieved, the
numerical evaluation of the solution remains challenging, and so an alternative route to efficient and stable numerical evaluation was taken. We obtain
a compact representation of the differential equation in this case by making $\dd \log$ choices for all master integrals except those in the
complicated sectors. For the problematic sectors, we find choices in which the differential equation is at most quadratic in $\eps$. This form is
compatible with rational reconstruction over finite fields using optimised IBP relations in the same fashion as the other topologies and with easily
manageable computation times. By using a set of linearly independent $\dd \log$ and non-logarithmic one-forms we express the DEs in a compact form where the number of non-logarithmic structures is minimised.

We evaluate the master integrals numerically by solving the corresponding DEs by means of \textsc{DiffExp}~\cite{Hidding:2020ytt}, a
\textsc{Mathematica} implementation of the method of generalised power series expansions~\cite{Francesco:2019yqt}.  We focus our analysis on the
physical phase-space region relevant for $pp \to t\bar{t}$+jet, the $s_{45}$ channel.  We obtain numerical boundary values in this region with
\textsc{AMFlow}~\cite{Liu:2022chg}, which implements the auxiliary mass flow method~\cite{Liu:2017jxz,Liu:2021wks,Liu:2022tji}.  Interestingly, we
observe that the master integrals related to the problematic features (that is, the nested square root and the elliptic curve) are non-zero only
starting from order $\eps^4$.  We perform a number of checks that the numerical evaluation in the $s_{45}$ channel is reliable and computationally
feasible. We find that the quadratic $\epsilon$-dependence of the DEs for topology ${\rm PB}_B$ results in an increase of the evaluation time of the method
with respect to the $\epsilon$-factorised DEs for topologies ${\rm PB}_A$ and ${\rm PB}_C$. Nevertheless, given the level of optimisation of our solution,
we find the performance analysis promising for future phenomenology applications.

There are still a number of issues to be addressed before the method can be applied in the context of amplitude computations. Proceeding without a
representation of the integrals in terms of a basis of special functions order by order in $\eps$ means that the poles could not be removed analytically. The inability to
perform an expansion in four dimensions may also mean that certain simplifications in the amplitude are not observed. Nevertheless, a strategy of
performing the integration-by-parts reduction to master integrals numerically over a rationalised phase space with modular arithmetic appears to be
achievable without major new technological developments. While combining this with an optimised strategy for the evaluation of the master integrals over
the full phase space is left for future work, we remark that the application of the generalised series expansion method for phenomenological studies
has been successful for other processes~\cite{Becchetti:2020wof,Armadillo:2022bgm,Bonciani:2021zzf,Becchetti:2023yat}.
It would also be interesting to continue the search for $\eps$-factorised DEs for ${\rm PB}_B$ following the latest line of research in this area,
which could potentially lead to even more efficient numerical evaluations. 
This would however require some substantial new developments in the available theoretical tools.

Our work paves the way to the analytic computation of the two-loop amplitudes for $pp\to t\bar{t}$+jet in the leading colour approximation, the main
bottleneck towards obtaining predictions for this important process at NNLO in QCD.

\acknowledgments

We are grateful to Heribertus Bayu Hartanto, Colomba Brancaccio, Ekta Chaubey and Christoph Dlapa for many enlightening discussions.
We also thank Colomba Brancaccio and Xuhang Jiang for helpful comments on this paper.
This project has received funding from the European Union's Horizon Europe research and innovation programme under the Marie Skłodowska-Curie grant agreement No.~101105486,
and ERC Starting Grant No.~101040760 \emph{FFHiggsTop}. This work has received funding from the Italian Ministry of Universities and Research through FARE
grant R207777C4R. This research was supported in part by the Swiss National Science Foundation (SNF) under contract 200021\_212729. SB has been partially supported by the Italian Ministry of Universities and Research (MUR) through grant PRIN 2022BCXSW9.

\appendix

\section{Maximal-cut analysis of the elliptic sector}
\label{sec:CutBaikov}

In this section, we analyse the maximal cut of the following scalar integral in the sector $321B$ of topology ${\rm PB}_B$ (see sec.~\ref{sec:321B} and fig.~\ref{fig:321B}):
\begin{align}
J = I^{({\rm PB}_B),0,0,0}_{1,1,0,1,1,1,0,1} \,.
\end{align}
This is related to the basis integral $\mathcal{I}_{{\rm PB}_B,35}$ through eq.~\eqref{eq:321B}.
The maximal cut is solution to the homogeneous DEs~\cite{Primo:2016ebd}, and therefore contains precious information about the analytic structure of the full solution.
We adopt the Baikov parametrisation~\cite{Baikov:1996iu,Frellesvig:2017aai}, and show that the maximal cut features the square root of a \mbox{degree-4} polynomial. 
The latter defines an elliptic curve, whose periods (suitably normalised) are solutions to the Picard-Fuchs operator of  $\mathcal{I}_{{\rm PB}_B,35}$ discussed in sec.~\ref{sec:321B}.

\begin{figure}[h]
\centering
\includegraphics[scale=0.5]{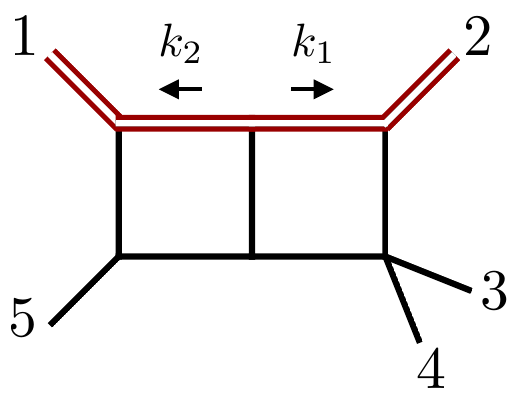}
\caption{Four-point double-box super-sector of the elliptic sector in topology ${\rm PB}_B$.}
\label{fig:dbTopoB}
\end{figure}

Instead of starting from the Baikov parametrisation of the pentagon-box top sector, we view this sector as a sub-sector of the four-point double-box shown in fig.~\ref{fig:dbTopoB}.
This allows us to get rid of two Baikov integration variables, corresponding to irreducible scalar products of the pentagon-box.
Furthermore, we have one fewer integration variable by adopting a loop-by-loop Baikov parametrisation~\cite{Frellesvig:2017aai}, starting from the $k_2$ loop.\footnote{Starting from the $k_1$ loop would lead to an equivalent result.}
Neglecting the overall kinematic-independent factors and the integration domain, our parametrisation is given by
\begin{align}
J \propto \left[{\rm det} \, G(p_1,p_2,p_5)\right]^{\eps} \int & \frac{\dd z_1 \, \dd z_2 \, \dd z_4 \, \dd z_5 \, \dd z_6 \, \dd z_8}{z_1 \, z_2 \, z_4 \, z_5 \, z_6 \, z_8} \, \dd z_7 \, \dd z_9 \times \nonumber \\
& \left[{\rm det} \, G(k_1,p_1,p_2,p_5)\right]^{-\frac{1}{2}-\eps} \left[{\rm det} \, G(k_1,p_1,p_5)\right]^{\eps} \left[{\rm det} \, G(k_2,k_1,p_1,p_5)\right]^{-\frac{1}{2}-\eps} \,,
\end{align}
where we recall that $G$ is the Gram matrix, i.e.\ $G_{ij}(v_1,\ldots,v_n) = v_i\cdot v_j $ for $i,j=1,\ldots,n$, and the scalar products $k_i \cdot k_j$ and $k_i \cdot p_j$ are understood as written in terms of inverse propagators $z_i$. 
The maximal cut, namely the residue of the integrand at $z_i = 0$ for $i=1,2,4,5,6,8$ in $d=4$ dimensions is given by
\begin{align} \label{eq:MaxCutJ}
{\rm MaxCut} \left[ J  \right] \bigl|_{\eps = 0} \propto \int \frac{\dd z_7 \, \dd z_9}{\sqrt{f_1(z_9) \, f_2(z_9) \, f_3(z_7,z_9) }} \,,
\end{align}
where the $f_i$'s are irreducible polynomials. In particular, $f_1(z_9) = z_9+m_t^2$, $f_2(z_9)$ is a degree-2 polynomial in $z_9$, and $f_3(z_7,z_9)$ is a polynomial of degree 1 in $z_9$ and 2 in $z_7$.
The dependence on the kinematic invariants is omitted.

We now wish to rewrite the integrand on the maximal cut as a linear combination of $\dd \log$ forms.
The coefficients of the $\dd \log$ forms are called leading singularities.
If this is possible, the integral ---~normalised so that its leading singularities are constant~--- is expected to satisfy canonical DEs~\cite{Arkani-Hamed:2010pyv}. 
For example, in the massless case (i.e., we set $m_t^2 = 0$ under the integral sign), we can write\footnote{We omit the exterior product between the differential forms to simplify the discussion.}
\begin{align}
{\rm MaxCut} \left[ J_{\rm massless}  \right] \bigl|_{\eps = 0} \propto \frac{1}{d_{12} d_{15}} \int \dd \log\left(z_7-2 d_{15}\right) \, \dd \log\left(1-\frac{2 d_{12} d_{15}}{z_9 (d_{15} - d_{34})}\right) \,,
\end{align}
which indicates that $d_{12} d_{15} \, J_{\rm massless}$ ---~modulo sub-sector corrections~--- is a good candidate for a canonical basis.
With a non-zero top-quark mass, however, only one differential form can be expressed as a $\dd \log$,
\begin{align} \label{eq:MaxCutJ2}
{\rm MaxCut} \left[ J  \right] \bigl|_{\eps = 0} \propto \int  \dd \log\left[\alpha(z_7,z_9)\right] \, \frac{\dd z_9}{\sqrt{\mathcal{P}(z_9)}} \,,
\end{align}
where $\alpha(z_7,z_9)$ is an irrelevant algebraic function, and $\mathcal{P}(z_9)$ is a degree-4 polynomial,
\begin{align}
\mathcal{P}(z_9) = \, & (z_9+m_t^2) (z_9 - 3 m_t^2) \bigl( \mathcal{P}_0 + \mathcal{P}_1 \, z_9 +  \mathcal{P}_2 \, z_9^2 \bigr) \,,
\end{align}
with
\begin{align}
\begin{aligned}
\mathcal{P}_0 & = 4 d_{12}^2 d_{15}^2 + 4 d_{12}  d_{15} \left( d_{15} + d_{34} \right) m_t^2 + \left( d_{15}^2 + 
4 d_{12} d_{34} + 2 d_{15} d_{34} - 3 d_{34}^2\right) m_t^4 + 2 d_{34} m_t^6  \,, \\
\mathcal{P}_1 & =  4 d_{12} d_{15} \left(d_{34} - d_{15}\right) - 2 \left(d_{15}^2 - 2 d_{12} d_{34} + 
 d_{34}^2 \right) m_t^2 \,, \\
\mathcal{P}_2 & =  \left(d_{15} - d_{34}\right)^2 - 2 d_{34} m_t^2 \,.
\end{aligned}
\end{align}
For generic values of the kinematic invariants, this degree-4 polynomial has four distinct roots, and thus defines an elliptic curve:
\begin{align} \label{eq:elliptic_curve}
y^2 = \mathcal{P}(x) \,.
\end{align}
The differential form which cannot be expressed as a $\dd \log$ in eq.~\eqref{eq:MaxCutJ2} is the holomorphic differential form of the first kind on this elliptic curve, which is one of the integration kernels defining the elliptic multiple polylogarithms~\cite{Broedel:2018qkq} (see also ref.~\cite{Gorges:2023zgv} for a similar example).
In the massless limit ($m_t^2 = 0$), instead, two of the roots degenerate, so that $\mathcal{P}(z_9)$ becomes a perfect square and the elliptic curve reduces to a genus-zero surface.
This is consistent with the fact that the massless pentagon-box integrals are of polylogarithmic type~\cite{Gehrmann:2015bfy,Papadopoulos:2015jft}.

The elliptic curve in eq.~\eqref{eq:elliptic_curve} contains useful information for putting the DEs in $\eps$-factorised form, and is tightly connected to the Picard-Fuchs operator of the integral $\mathcal{I}_{{\rm PB}_B,35}$ discussed in sec.~\ref{sec:321B}.
In particular, we expect that the Feynman integral $J$ normalised by a period of the elliptic curve in eq.~\eqref{eq:elliptic_curve} is a good candidate for a canonical integral~\cite{Gorges:2023zgv}.
While we leave this study for further work, we discuss here the connection with the Picard-Fuchs operator.

For this purpose, we must first spell out the periods of the elliptic curve.
The four roots are given by\footnote{In the $s_{45}$ channel we have that $e_1 < 0$, $e_4>0$, $\delta < 0$, while the real parts of $e_2$ and $e_3$ do not have fixed sign.}
\begin{align}
\begin{aligned}
& e_1 = - m_t^2 \,, \\
& e_2 = \frac{2 d_{12} d_{15} (d_{34}-d_{15}) - (d_{15}^2 + d_{34}^2 - 2 d_{12} d_{34})  m_t^2 + d_{34} \sqrt{\delta}}{{\rm det}\, G(p_2,p_1+p_5)} \,, \\
& e_3 = e_2 \bigl|_{\sqrt{\delta}\to -\sqrt{\delta}}  \,, \\
& e_4 = 3 m_t^2 \,,
\end{aligned}
\end{align}
where
\begin{align}
\delta = - 4 \, m_t^2 \, {\rm det} \, G(p_1,p_2,p_5) \,.
\end{align}
The periods can then be chosen as~\cite{Weinzierl:2022eaz}
\begin{align}
\psi_1 = \frac{4 \, {\rm K}(\kappa^2)}{\sqrt{(e_3-e_1)(e_4-e_2)}} \,, \qquad \quad
\psi_2 = \frac{4 \, \ii \, {\rm K}(1-\kappa^2)}{\sqrt{(e_3-e_1)(e_4-e_2)}} \,,
\end{align}
where ${\rm K}(x)$ is the elliptic integral of the first kind, and $\kappa$ is the modulus of the elliptic curve,
\begin{align}
\kappa^2 = \frac{(e_3-e_2) (e_4-e_1)}{(e_3-e_1)(e_4-e_2)} \,.
\end{align}
Note that $e_2$ and $e_3$ involve a square root, and thus the periods feature another nested square root, in addition to the one discussed in sec.~\ref{sec:421B}.
We find that ---~with a suitable algebraic normalisation~--- the periods $\psi_i$ are solutions to the Picard-Fuchs operator $L_{35}$ of the Feynman integral $\mathcal{I}_{{\rm PB}_B,35}$ (modulo sub-sectors and $\eps$-corrections).
Explicitly, we have that
\begin{align}
L_{35} \, \left[\frac{d_{15} (d_{12}+m_t^2) \, \psi_i}{\sqrt{(d_{15}-d_{34})^2-2 d_{34} m_t^2}} \right]  = 0 \,,
\end{align}
for $i =1,2$, where the terms in the square brackets are understood as evaluated on the univariate phase-space slice used to derive $L_{35}$.
The rational factors multiplying the periods follow from the chosen normalisation of $\mathcal{I}_{{\rm PB}_B,35}$ (see eq.~\eqref{eq:321B}), while the square root is the coefficient of the degree-$4$ monomial in the polynomial $\mathcal{P}$ defining the elliptic curve. 
With this, we have established that the elliptic integrals found in the solutions to the Picard-Fuchs operator $L_{35}$ in sec.~\ref{sec:321B} and thus in the homogeneous solution to the DEs are associated with the elliptic curve given in eq.~\eqref{eq:elliptic_curve}.

\section{All one-loop pentagon integrals}
\label{app:pentagons}

\begin{figure}[t!]
	\centering
	\begin{subfigure}{0.25\linewidth}
		\includegraphics[width=\linewidth]{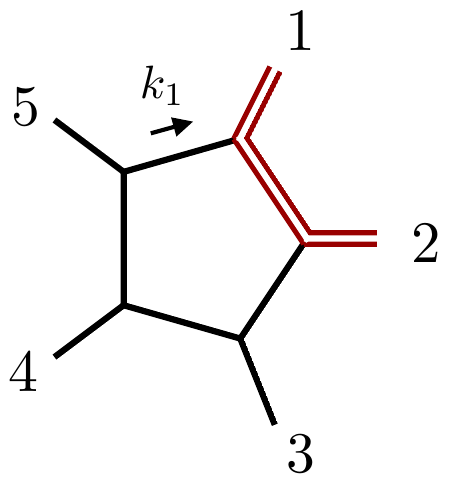}
		\caption{Topology ${\rm P}_A$.}
		\label{fig:PttjA}
	\end{subfigure} \hspace{0.2cm}
	\begin{subfigure}{0.25\linewidth}
		\includegraphics[width=\linewidth]{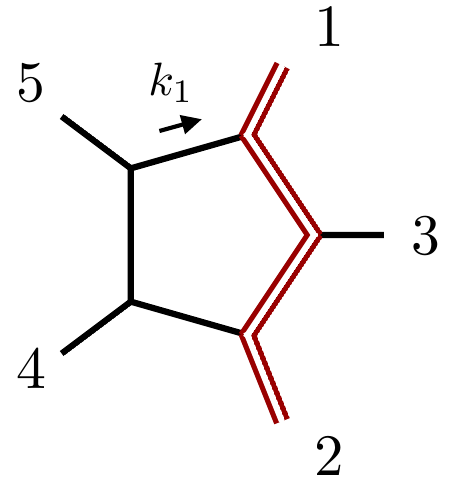}
		\caption{Topology ${\rm P}_B$.}
		\label{fig:PttjB}
	\end{subfigure} \\ \vspace{0.2cm}
	\begin{subfigure}{0.25\linewidth}
		\includegraphics[width=\linewidth]{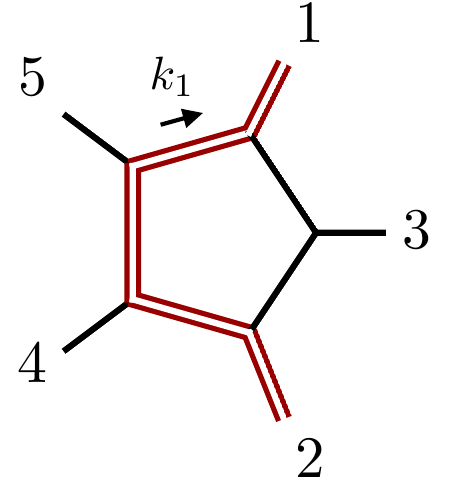}
		\caption{Topology ${\rm P}_C$.}
		\label{fig:PttjC}
	\end{subfigure} \hspace{0.2cm}
	\begin{subfigure}{0.25\linewidth}
		\includegraphics[width=\linewidth]{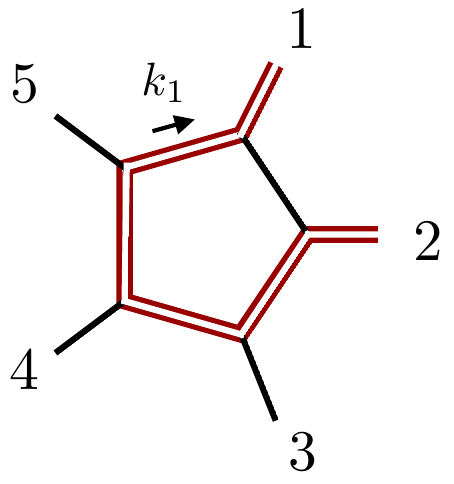}
		\caption{Topology ${\rm P}_D$.}
		\label{fig:PttjD}
	\end{subfigure}
        \caption{Graphical representation of the four one-loop pentagon topologies.}
	\label{fig:1Lpentagons}
\end{figure}

We update the representation of the differential equations for all one-loop
pentagon integrals contributing to $pp\to t\tb+$jet previously considered in
reference~\cite{Badger:2022mrb}. We present them in `$\dd \log$' form
using an alphabet which shares as much in common with the two-loop planar
integrals as possible. For completeness we include here the definitions of the integral families which are shown graphically in figure~\ref{fig:1Lpentagons}.
The integrals of the pentagon topology $F \in \{{\rm P}_A,{\rm P}_B,{\rm P}_C,{\rm P}_D\}$ have the form
\begin{equation}
  I_{a_1,a_2,a_3,a_4,a_5}^{\left(F\right)} = \int \mathcal{D}^{d} k_1 \,
  \frac{1}{D_{F,1}^{a_1}\cdots D_{F,5}^{a_5}}\,,
  \label{eq:t5def}
\end{equation}
where we recall that the integration measure $\mathcal{D}^{d} k_1$ is defined in eq.~\eqref{eq:measure}.
The inverse propagators $D_{F,i}$ are defined in table~\ref{tab:pentagon-propagators}.
\begin{table}[h!]
\begin{center}
\begin{tabular}{|c|c|c|c|c|}
\hline
& ${\rm P}_A$ & ${\rm P}_B$ & ${\rm P}_C$ & ${\rm P}_D$ \\
\hline
$D_{F,1}$ & $k_1^2$ & $k_1^2$ & $k_1^2-m_t^2$ & $ k_1^2-m_t^2$ \\
$D_{F,2}$ & $(k_1-p_1)^2-m_t^2$ & $ (k_1-p_1)^2-m_t^2 $ & $(k_1-p_1)^2$ & $(k_1-p_1)^2$  \\
$D_{F,3}$ & $(k_1-p_1-p_2)^2$ & $ (k_1-p_1-p_3)^2-m_t^2$ & $(k_1-p_1-p_3)^2$ & $(k_1-p_1-p_2)^2-m_t^2$ \\
$D_{F,4}$ & $(k_1+p_4+p_5)^2$ & $ (k_1+p_4+p_5)^2$ & $(k_1+p_4+p_5)^2-m_t^2$ & $(k_1+p_4+p_5)^2-m_t^2$ \\
$D_{F,5}$ & $(k_1+p_5)^2$ & $(k_1+p_5)^2$ & $(k_1+p_5)^2-m_t^2$ & $(k_1+p_5)^2-m_t^2$ \\
\hline
\end{tabular}
\end{center}
\caption{Inverse propagators $D_{F,i}$ of the pentagon topologies shown in figure~\ref{fig:1Lpentagons}.}
\label{tab:pentagon-propagators}
\end{table}

\section{Description of the ancillary files}
\label{sec:anc}

The ancillary files can be downloaded from ref.~\cite{ancillary}.
The symbols in the files are given in the notation of the article as follows:
\begin{align*}
\texttt{eps} & = \eps \,, &  \texttt{W[i]} & = W_i \,, \\
\texttt{dij} & = d_{ij} \,, & \texttt{j[F,a1,\ldots,a5]} & = I^{(F)}_{a_1, \ldots, a_5} \,,  \\
\texttt{mt2} & = m_t^2 \,, & \texttt{j[F,a1,\ldots,a11]} & = I^{(F),-a_{9},-a_{10},-a_{11}}_{a_1, \ldots, a_8} \,, \\
 \texttt{s[i,j]} & = s_{ij} \,, &   \texttt{dlog[W[i]]} & = {\rm d} \log\left( W_i \right) \,,  \\
 \texttt{s[i]} & = p_i^2 \,, & \texttt{of[i]} & = \omega_i \,, \\
 \texttt{pi} & = p_i \,, & \texttt{GramDet[{P1,\ldots,Pn}]} & = {\rm det} \, G(P_1,\ldots,P_n) \,,  & \\
 \texttt{ki} & = k_i \,, & \texttt{trp[i1,\ldots,in]} & = {\rm tr}_+\left[\slashed{p}_{i_1} \ldots \slashed{p}_{i_n} \right] \,, \\
\texttt{p[i,j]} & = p_{ij} \,, &  \texttt{trm[i1,\ldots,in]} & = {\rm tr}_-\left[\slashed{p}_{i_1} \ldots \slashed{p}_{i_n} \right] \,, \\
\texttt{spAB[i,\ldots,j]} & =\langle i|\ldots|j] \,, &  \texttt{sqrtratio[a,b]} & = \frac{a+b}{a-b} \,. \\
\end{align*}

The following files give global definitions of the square roots, the alphabet letters and the physical scattering region under consideration.
\begin{itemize}
\item \texttt{square\_roots.m} -- definition of the square roots in the format
\begin{equation*}
\texttt{\{ \ldots, S -> Sqrt[expr] , \ldots \}}\,,
\end{equation*}
where $\texttt{S}$ is the square-root label,
\begin{alignat*}{8}
\texttt{tr5} & = {\rm tr}_5 , \quad
\texttt{beta12} && = \beta_{12} , \quad
\texttt{beta34} && = \beta_{34} , \quad 
\texttt{beta45} && = \beta_{45} , \quad
\texttt{Delta31} && = \Delta_{3,1}, \\
\texttt{Delta32} & = \Delta_{3,2}, \quad
\texttt{Delta33} && = \Delta_{3,3}, \quad
\texttt{Delta34} && = \Delta_{3,4}, \quad
 \texttt{Lambda1} && = \Lambda_1 , \quad
 \texttt{Lambda2} && = \Lambda_2 , \\
\texttt{Lambda3} & = \Lambda_3 , \quad
\texttt{Lambda4} && = \Lambda_4 , \quad
\texttt{Lambda5} && = \Lambda_5 , \quad
\texttt{Lambda6} && = \Lambda_6 , &&
\end{alignat*}
and \texttt{expr} is a rational function of the invariants $\vec{x}$.
See sec.~\ref{sec:setup} for a representation in terms of Gram and Cayley determinants.
 
\item \texttt{alphabet.m} -- definition of the alphabet letters $W_i$ in the format
\begin{equation*}
\texttt{\{\ldots, W[i] -> expr, \ldots \}}\,,
\end{equation*}
where \texttt{expr} is given in terms of invariants $\vec{x}$, $s_{ij}$ variables, Gram determinants, spinor-helicity chains, and gamma-matrix traces.

\item \texttt{alphabet\_dij.m} -- definition of the alphabet letters \texttt{W[i]} in the same format as in \texttt{alphabet.m}, but with \texttt{expr} given in terms of invariants $\vec{x}$ and square roots.

\item \texttt{alphabet\_charges.m} -- charges of the letters in the format 
\begin{equation*}
\texttt{\{\ldots, C -> \{W[i1],W[i2],\ldots\},\ldots\}} \,,
\end{equation*}
meaning that the letters $\{W_{i_1}, W_{i_2}, \ldots\}$ are odd with respect to the square-root monomial \texttt{C}.
The letters with \texttt{C}$\ =1$ are rational, and thus even with respect to all square roots.
See the introduction of sec.~\ref{sec:deqs} for the notion of charge.

\item \texttt{s45-channel.m} -- the inequalities in the $d_{ij}$ variables which define the $s_{45}$ channel (see sec.~\ref{sec:s45channel}).

\end{itemize}

For each integral topology \texttt{<fam>} studied in this work there is a folder with the same name containing the definition of the propagators and of the integral basis, the differential equations, and the boundary values.
The integral topologies are denoted by 
\begin{gather*}
\texttt{PttjA} = {\rm P}_A \,, \qquad
\texttt{PttjB} = {\rm P}_B \,, \qquad
\texttt{PttjC} = {\rm P}_C \,, \qquad
\texttt{PttjD} = {\rm P}_D \,, \\
\texttt{PBttjA} = {\rm PB}_A \,, \qquad
\texttt{PBttjB} = {\rm PB}_B \,, \qquad
\texttt{PBttjC} = {\rm PB}_C \,.
\end{gather*}
The folder of the family \texttt{<fam>} contains the following files.
\begin{itemize}
\item \texttt{<fam>\_propagators.m} -- the inverse propagators $D_{F,i}$.

\item \texttt{<fam>\_basis\_definitions.m} -- the definition of the (square-root free) integral basis in the format
\begin{align*}
\texttt{\{\ldots, mi[<fam>,i] -> expr, \ldots \}}\,,
\end{align*}
where \texttt{mi[<fam>,i]} denotes the $i$-th master integral of the topology \texttt{<fam>}, and \texttt{expr} is a linear combination of scalar integrals with coefficients given by rational functions of the invariants $\vec{x}$ and~$\eps$.

\item \texttt{<fam>\_basis\_norm.m} -- the square-root normalisations that should be applied in addition to the basis integrals. More explicitly, the $i$-th master integral in \texttt{<fam>\_basis\_definitions.m} has to be multiplied by the $i$-th entry of \texttt{<fam>\_basis\_norm.m}. The separation between rational expressions and square roots is motivated in sec.~\ref{sec:deqs}.

\item \texttt{<fam>\_de\_dlogs.m} (or \texttt{PBttjB\_de\_one-forms.m} where the $\dd \log$-representation is not possible) -- the connection matrices of the differential equations for the master integrals expressed in terms of $\dd \log$'s and, for ${\rm PB}_B$, one-forms $\omega_i$.

\item \texttt{<fam>\_boundary\_values\_s45.m} -- phase-space point and numerical values of the master integrals in the $s_{45}$ scattering region with at least 32-digit precision. The format is \texttt{\{X,values\}}. The first entry, \texttt{X}, gives the boundary point $\vec{x}_0$ defined in eq.~\eqref{eq:bound_point}, in the format
\begin{align*}
\texttt{<|d12->2, d23->1, d34->-1, d45->5, d15 ->-2, mt2->1|>}\,.
\end{align*}
The second entry, \texttt{values}, is a $n_{\rm MI} \times 5$ array, where $n_{\rm MI}$ is the number of master integrals of the family, and $5$ is the number of orders in $\eps$ (from $\eps^0$ to $\eps^4$).
The entry $(i,j)$ of \texttt{values} gives the value of the $\mathcal{O}\left(\eps^{j-1}\right)$ coefficient of the $i$-th master integral.

\end{itemize}
The folder for topology ${\rm PB}_B$ contains two additional files, due to the fact that the differential equations cannot be expressed in terms of $\dd \log$'s only.
\begin{itemize}

\item \texttt{PBttjB\_one-forms\_definitions.m} -- definition of the one-forms $\omega_i$ in the format
\begin{align*}
\texttt{\{\ldots, of[i] -> \{r1,r2,\ldots,r6\}, \ldots\}}\,,
\end{align*}
meaning that 
\begin{align*}
\omega_i(\vec{x}) = \sum_{j=1}^6 r_j(\vec{x}) \, \dd x_j \,,
\end{align*}
where $r_j = \ $\texttt{rj} are functions of the invariants $\vec{x}$ and of the square roots. 

\item \texttt{PBttjB\_one-forms\_charges.m} -- charges of the one-forms in the format
\begin{align*}
\texttt{\{\ldots,C -> \{of[i1],of[i2],\ldots\},\ldots\}}\,,
\end{align*}
meaning that the one-forms $\{\omega_{i_1}$, $\omega_{i_2},\ldots\}$ are odd with respect to the square-root monomial~\texttt{C}.
The one-forms with \texttt{C}$\ =1$ are rational, and thus even with respect to all square roots.
\end{itemize}

The folder \texttt{benchmarks/} contains benchmark values of all basis integrals.
\begin{itemize}
\item \texttt{point\_x1\_s45.m} -- phase-space point $\vec{x}_1$ in the $s_{45}$ channel defined in eq.~\eqref{eq:x1}, in the format
\begin{align} \label{eq:PSformat}
\texttt{\{d12->\#val, d23->\#val, d34->\#val, d45->\#val, d15 ->\#val, mt2->\#val\}}\,,
\end{align}
where \texttt{\#val} denotes a numerical value.

\item \texttt{<fam>\_values\_x1.m} -- values of the basis integrals of family \texttt{<fam>} at the phase-space point $\vec{x}_1$ with at least 32-digit precision. The format is the same as in \texttt{<fam>\_boundary\_values\_s45.m}.

\end{itemize}

A \textsc{Mathematica} script, \texttt{DiffExp\_run.wl}, is provided to evaluate numerically the basis integrals through \textsc{DiffExp}~\cite{Hidding:2020ytt}.
The path to \texttt{DiffExp.m} has to be specified through the variable \texttt{PathToDiffExp}.
The script may be used via the command line as
\begin{align*}
\texttt{math -script DiffExp\_run.wl [-family <fam>] [-target <file>] [-storepiecewise]}
\end{align*}
The file indicated to be the target phase-space point is expected in the format shown in eq.~\eqref{eq:PSformat}.
If the option \texttt{-storepiecewise} is given, the script also stores the analytic expression of the generalised power series solution to the differential equations. 
Otherwise, only the numerical values of the basis integrals at the target point are saved, in the \texttt{benchmarks/} folder.
The integration of the differential equations occurs along a straight path connecting the target point to the boundary point $\vec{x}_0$ in eq.~\eqref{eq:bound_point}.
If the path leaves the $s_{45}$ channel, the evaluation is aborted.
This may occur even if the target point is in the $s_{45}$ channel. 
In such a case, one should either choose a different starting point for the integration path, or study the analytic continuation from the $s_{45}$ channel to the region of interest.
The latter is instead mandatory if the target point is outside of the $s_{45}$ channel.
The accuracy goal is hard-coded to 16 digits.
If the target point is $\vec{x}_1$, the evaluations are compared against the provided benchmark values.

\bibliographystyle{JHEP}
\bibliography{ppttj_PTopo}

\end{document}